\pdfoutput=1
\PassOptionsToPackage{accepted}{servecast-style}
\documentclass{article}
\usepackage{microtype,graphicx,booktabs,amsmath,amssymb,array}
\usepackage{multirow}
\usepackage{tikz}
\usetikzlibrary{arrows.meta,positioning,calc}
\usepackage{fontawesome5}
\newcommand{\figfont}{\fontfamily{phv}\selectfont}

\usepackage{colortbl}

\usepackage{enumitem}

\definecolor{spblue}{HTML}{1F6FB4}\definecolor{vidred}{HTML}{C0392B}
\definecolor{ink}{HTML}{263238}
\definecolor{pfgreen}{HTML}{2E9E5B}
\definecolor{dcorange}{HTML}{F39C12}
\definecolor{kvblue}{HTML}{4A90D9}
\definecolor{evict}{HTML}{8E44AD}
\definecolor{offfill}{HTML}{FFF3D1}
\definecolor{offline}{HTML}{D68910}
\definecolor{onfill}{HTML}{E4F0FB}
\definecolor{gpufill}{HTML}{EAF3FC}
\colorlet{mixcol}{pfgreen!50!dcorange!80!black}
\newcommand{\winS}[1]{\cellcolor{spblue!14}\textbf{#1}}
\newcommand{\winV}[1]{\cellcolor{vidred!13}\textbf{#1}}
\usepackage{xurl}
\usepackage{hyperref}
\usepackage{servecast-style}

\newcommand{\rqanswer}[2]{\par\smallskip\noindent\fbox{\parbox{\dimexpr\linewidth-2\fboxsep-2\fboxrule\relax}{%
\textbf{#1}\par\smallskip #2}}\par\smallskip}

\newcommand{\arxivsystemname}{ServeCast}

\newcommand{\arxivtitle}{ServeCast: Predicting LLM Serving Throughput and Latency Across Models from Microbenchmarks}

\newcommand{\arxivauthorblock}{%
\begin{castauthorlist}
\castauthor{Emilio Incerto}{imt}
\castauthor{Lishan Yang}{gmu}
\castauthor{Yu Sun}{gmu}
\castauthor{Mirco Tribastone}{imt}
\end{castauthorlist}
\castaffiliation{imt}{IMT School for Advanced Studies Lucca, Italy}
\castaffiliation{gmu}{George Mason University, USA}
\castcorrespondingauthor{Emilio Incerto}{emilio.incerto@imtlucca.it}
}

\makeatletter
\renewcommand{\Notice@String}{Preprint.}
\let\arxiv@addtocurcol\@addtocurcol
\usepackage{flafter}
\let\arxiv@flafter@addtocurcol\@addtocurcol
\let\@addtocurcol\arxiv@addtocurcol
\newcommand{\arxivappendixfloats}{\let\@addtocurcol\arxiv@flafter@addtocurcol}
\makeatother
\usepackage{longtable,placeins}
\renewcommand{\winS}[1]{\cellcolor{pfgreen!17}\textbf{#1}}
\renewcommand{\winV}[1]{\cellcolor{dcorange!24}\textbf{#1}}
\newcommand{\arxivdiag}[1]{\cellcolor{kvblue!15}\textbf{#1}}
\hypersetup{pdftitle={\arxivtitle}, pdfsubject={}, pdfkeywords={}}

\usepackage{balance}
\casttitlerunning{ServeCast}
\begin{document}
\twocolumn[
\casttitle{\arxivtitle}
\arxivauthorblock
\castkeywords{LLM serving, performance prediction, continuous batching, cross-model prediction}
\vskip 0.3in
\begin{abstract}
Deciding how to deploy an LLM means predicting its throughput and latency under varying load. Benchmarking every candidate model on every GPU is expensive, and prediction is complicated by continuous batching and by request contexts that evolve. We introduce ServeCast, a performance model that follows requests with variable prompt and output lengths through admission, prefill, and decode under limited GPU memory. Its explicit parameters describe the LLM architecture and the clock and memory bandwidth of the GPU, and three timing coefficients are calibrated with microbenchmarks. To predict how a different LLM is served on the same GPU, ServeCast may transfer the performance model. The transfer updates the explicit parameters and keeps the timing coefficients fitted on a different LLM, so no re-profiling is needed. Our extensive numerical evaluation spans three GPUs, six LLMs, and three trace-derived workloads, comparing ServeCast with Vidur, a widely used simulator. When both are calibrated on the target, the aggregate mean absolute percentage error (MAPE) of ServeCast is 6.4\% for mean request throughput, 11.9\% for time to first token (TTFT), and 7.5\% for time per output token (TPOT), against 8.9\%, 25.0\%, and 8.6\% for Vidur, respectively. Under cross-model transfer, throughput and TPOT errors are comparable to those of Vidur, which requires calibration for each model/GPU pair, while ServeCast's aggregate TTFT predictions are still more accurate.
\end{abstract}
]
\printAffiliationsAndNotice{}
\section{Introduction}
\label{sec:intro}

An LLM inference provider must select model and GPU configurations that meet the service level objectives it targets. Performance modeling is central to planning resources and to weighing trade-offs. In batching, for instance, processing work from several requests together on the GPU can improve throughput \cite{sarathi-osdi24}; yet, admitting more requests raises their aggregate memory demand and can raise latency \cite{slim26}.

\noindent {\bf Challenges.} First, every change in models, GPUs, or serving settings enlarges the set of candidate configurations to analyze~\cite{vidur24}; measurements  require representative workloads run at several intensities.
Second, the performance optimizations of LLM serving make inference dynamics complex. Under continuous batching, for example, batch composition sets iteration duration, and the progress and completion of requests set the next batches~\cite{orca-osdi22}. Prompt and output lengths shape both the work and the memory a request needs. When memory runs short, requests can be suspended, and recomputation is then required~\cite{vllm-sosp23,vllm-chunked-prefill-docs}. A prediction of serving performance must therefore couple workload characteristics with scheduling under memory constraints and with GPU execution time.

\noindent {\bf Limitations of the state of the art.} The approaches available today trade fidelity against cost. Vidur simulates scheduling and the progression of requests with operator timing estimates from profiling~\cite{vidur24}. These estimates are specific to one model/GPU configuration, so every new pair needs a profiling run of its own. Analytical and operator-level models, LIFE and LLMCompass among them, avoid this cost~\cite{life25,llmcompass-isca24}. Their latency estimates hold for a fixed inference configuration, however, and leave out the evolving batches and the memory contention of a serving system. SLIM and the queueing model of Ramani and Tantawi capture serving load~\cite{slim26,queueing-autoscaling26}. Because both describe requests by their mean lengths, they may overlook the length variability behind memory pressure, which in turn affects prediction accuracy.

\noindent{\bf {Key insights and contributions.}}
ServeCast, the performance model presented here, predicts the throughput and latency of LLM serving from explicit architecture inputs together with calibrated timing coefficients. It supports cross-model transfer: To predict the performance of another LLM on the same GPU, ServeCast reuses the timing coefficients and updates the architecture inputs, so the target LLM needs no new profiling.

ServeCast is built from \emph{two ingredients}.

The first is a request-level description of LLM serving dynamics, which captures how requests and their prompts progress through the execution stages. Prompt and output lengths follow a probability distribution that represents their variation and their correlation. With these choices, the model can represent batches that mix requests of different lengths at different execution stages, and it can determine how those requests require affects admission, suspension, and reconstruction.

The second ingredient estimates the iteration duration of a batch composition. The underlying model accounts for GPU compute and memory traffic with relations for the dense layers, the vocabulary projection, and the attention mechanism. Its explicit quantities come from both the LLM and the GPU architectures--- for instance, the numbers of layers and query heads, the GPU clock frequency, and the memory bandwidth. GPU microbenchmarks calibrate three timing coefficients, one for a compute efficiency term, similar in spirit to earlier execution-time models such as LIFE~\cite{life25} and GenZ~\cite{genz24}, and two for residual terms.

The separation of architectural inputs from fitted timing coefficients makes cross-model transfer between LLMs on the same GPU possible for predicting performance metrics such as the request throughput, the time to first token (TTFT), and the time per output token (TPOT).

We evaluate ServeCast through three research questions.
\begin{itemize}[nosep]
\item \textbf{RQ1: Prediction accuracy.} How accurate are the throughput, TTFT, and TPOT predictions of ServeCast when it is calibrated on the target model--GPU pair?

\item \textbf{RQ2: Cross-model transfer.} With timing coefficients calibrated on a different LLM on the same GPU and no timing profiling of the target, how accurately does ServeCast predict the target LLM's performance?

\item \textbf{RQ3: Profiling and prediction time.} How much GPU time does profiling for ServeCast consume, and how long does ServeCast need to produce its predictions?
\end{itemize}
We compare ServeCast with measured vLLM performance on three GPUs, six LLMs, and three trace-derived workloads, with target-profiled Vidur as the baseline. Under target-specific calibration, ServeCast has lower aggregate errors than Vidur on all three metrics, with a TTFT MAPE of 11.9\% against 25.0\% for Vidur. When timing coefficients are reused across LLMs on the same GPU, the target model needs no timing profiling, and the aggregate TTFT error stays below Vidur's while the throughput and TPOT errors remain comparable. The mean prediction time per configuration is 13.1 seconds for ServeCast and 25.9 seconds for Vidur.

ServeCast is open source. Its code is available at \url{https://github.com/servecast/servecast}, and its documentation at \url{https://servecast.github.io/servecast/}.

\section{Background}
\label{sec:bg}
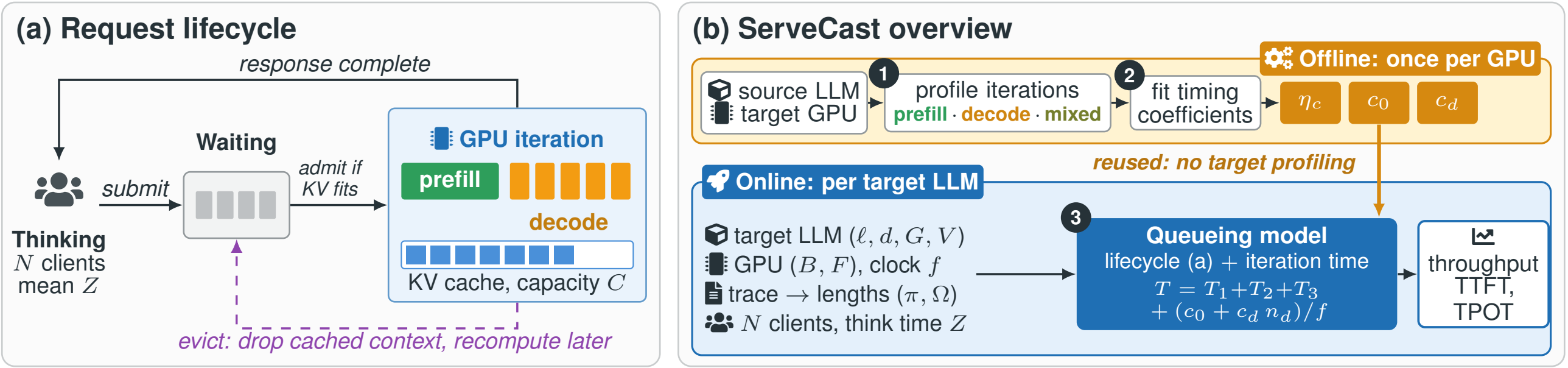
\begin{figure*}[t]\centering
\begin{tikzpicture}[font=\figfont\scriptsize, ink,
  arr/.style={-{Latex[length=1.5mm,width=1.3mm]}, line width=0.65pt, ink},
  panel/.style={draw=ink!25, rounded corners=4pt, line width=0.5pt, fill=ink!2},
  card/.style={draw=ink!50, fill=white, rounded corners=2pt, line width=0.5pt, align=center, inner sep=2.2pt},
  lab/.style={align=center, inner sep=0.8pt, font=\figfont\scriptsize\itshape},
  ptitle/.style={font=\figfont\footnotesize\bfseries, anchor=north west, inner sep=0pt},
  tag/.style={rounded corners=1.5pt, inner sep=1.5pt, text=white, font=\figfont\scriptsize\bfseries, anchor=west},
  stepn/.style={circle, fill=ink, text=white, inner sep=0pt, minimum size=3.3mm, font=\figfont\tiny\bfseries},
  coef/.style={fill=offline, text=white, rounded corners=1.5pt, minimum width=6.6mm, minimum height=4.6mm, inner sep=1pt},
]
\draw[panel] (0,0) rectangle (7.15,3.95);
\node[ptitle] at (0.15,3.8) {(a) Request lifecycle};

\node[font=\large, ink!85] (users) at (0.62,1.92) {\faUsers};
\node[align=center, below=0.2mm of users] (ulab) {\textbf{Thinking}\\[-1pt]$N$ clients\\[-1pt]mean $Z$};

\node[card, fill=ink!6, minimum width=11.5mm, minimum height=7.2mm] (queue) at (2.55,1.75) {};
\foreach \i in {0,...,3} \fill[ink!30, rounded corners=0.5pt] ($(queue.west)+(1.4mm+\i*2.3mm,-1.5mm)$) rectangle ++(1.8mm,3.0mm);
\node[above=0.5mm of queue, font=\figfont\scriptsize\bfseries] {Waiting};

\node[card, draw=spblue!70, fill=gpufill, minimum width=28mm, minimum height=21mm] (gpu) at (5.6,1.75) {};
\node[anchor=north, font=\figfont\scriptsize\bfseries, text=spblue] at ($(gpu.north)+(0,-0.9mm)$) {\faMicrochip~GPU iteration};
\fill[pfgreen, rounded corners=1pt] ($(gpu.center)+(-12.6mm,0.6mm)$) rectangle ++(10.6mm,4.0mm);
\node[white, font=\figfont\scriptsize\bfseries] at ($(gpu.center)+(-7.3mm,2.6mm)$) {prefill};
\foreach \i in {0,...,4} \fill[dcorange, rounded corners=0.7pt] ($(gpu.center)+(-0.8mm+\i*2.75mm,0.6mm)$) rectangle ++(2.1mm,4.0mm);
\node[anchor=north, text=dcorange!80!black, font=\figfont\scriptsize\bfseries] at ($(gpu.center)+(5.6mm,0.3mm)$) {decode};
\draw[draw=kvblue, rounded corners=1pt, fill=white] ($(gpu.center)+(-12.6mm,-6.9mm)$) rectangle ++(25.2mm,3.0mm);
\foreach \i in {0,...,6} \fill[kvblue] ($(gpu.center)+(-12.1mm+\i*2.68mm,-6.4mm)$) rectangle ++(2.18mm,2.0mm);
\node[font=\figfont\scriptsize] at ($(gpu.center)+(0,-8.6mm)$) {KV cache, capacity $C$};

\draw[arr] ($(users.east |- queue.west)+(0.05,0)$) -- node[lab, above=0.5mm, pos=0.42]{submit} (queue.west);
\draw[arr] (queue.east) -- node[lab, above=0.6mm, pos=0.4, font=\figfont\fontsize{6}{6.6}\selectfont\itshape]{admit if\\KV fits} (gpu.west);
\draw[arr] (gpu.north) -- ++(0,0.3) -| node[lab, above, pos=0.2]{response complete} (users.north);
\draw[arr, evict, dashed] (gpu.south) -- ++(0,-0.28) -| node[lab, below, pos=0.22, text=evict]{evict: drop cached context, recompute later} (queue.south);

\draw[panel] (7.35,0) rectangle (16.98,3.95);
\node[ptitle] at (7.5,3.8) {(b) ServeCast overview};

\draw[draw=offline, rounded corners=3pt, fill=offfill, line width=0.6pt] (7.5,2.42) rectangle (16.83,3.34);
\node[tag, fill=offline, anchor=east] at (16.73,3.34) {\faCogs~Offline: once per GPU};
\node[card, text width=16.5mm] (src) at (8.5,2.86) {\faCube~source LLM\\[-1pt]\faMicrochip~target GPU};
\node[card, text width=23mm] (probe) at (10.82,2.86) {profile iterations\\[-1pt]{\fontsize{6}{6.6}\selectfont{\color{pfgreen}\bfseries prefill}\,$\cdot$\,{\color{dcorange!85!black}\bfseries decode}\,$\cdot$\,{\color{mixcol}\bfseries mixed}}};
\node[card, text width=12.6mm] (fit) at (12.97,2.86) {fit timing\\[-1pt]coefficients};
\node[coef] (c1) at (14.22,2.86) {$\eta_c$};
\node[coef, right=0.7mm of c1] (c2) {$c_0$};
\node[coef, right=0.7mm of c2] (c3) {$c_d$};
\node[stepn] at (probe.north west) {1};
\node[stepn] at (fit.north west) {2};
\draw[arr] (src) -- (probe);
\draw[arr] (probe) -- (fit);
\draw[arr] (fit) -- (c1);

\draw[draw=spblue, rounded corners=3pt, fill=onfill, line width=0.6pt] (7.5,0.12) rectangle (16.83,2.0);
\node[tag, fill=spblue] at (7.6,2.0) {\faRocket~Online: per target LLM};
\node[anchor=west, align=left, inner sep=1pt] (in) at (7.6,0.95) {%
  \faCube~target LLM ($\ell, d, G, V$)\\[1pt]
  \faMicrochip~GPU ($B, F$), clock $f$\\[1pt]
  \faFile*~trace $\rightarrow$ lengths ($\pi, \Omega$)\\[1pt]
  \faUsers~$N$ clients, think time $Z$};
\node[fill=spblue, text=white, rounded corners=2.5pt, align=center, inner sep=2.6pt, text width=33mm, font=\figfont\fontsize{6.5}{7.3}\selectfont] (qm) at (13.42,1.0) {%
  {\scriptsize\bfseries Queueing model}\\[0.3pt]
  lifecycle (a) $+$ iteration time\\[1pt]
  \mbox{$T = T_1{+}T_2{+}T_3$}\\[-0.5pt]\mbox{${}+(c_0 + c_d\,n_d)/f$}};
\node[stepn] at (qm.north west) {3};
\node[card, draw=spblue, line width=0.7pt, text width=12.6mm] (out) at (16.1,1.0) {\faChartLine\\[0pt]throughput\\[-1pt]TTFT, TPOT};
\draw[arr] ($(in.east |- qm.west)+(0.05,0)$) -- (qm.west);
\draw[arr] (qm.east) -- (out.west);
\draw[arr, offline, line width=1.2pt] (c2.south) -- (c2.south |- qm.north);
\node[anchor=east, text=offline!85!black, font=\figfont\scriptsize\bfseries\itshape] at ($(c2.south)+(-0.14,-0.43)$) {\fontsize{6.5}{7}\selectfont reused: no target profiling};
\end{tikzpicture}
\vspace{-9pt}
\caption{Brief overview of ServeCast, with (a) the request lifecycle and (b) the workflow.}\label{fig:overview}\end{figure*}

\noindent\textbf{LLM inference.}
A decoder-only LLM generates text token by token through attention and feed-forward transformer blocks, then projects onto the token vocabulary~\cite{vllm-sosp23}. Each request has two phases. \emph{Prefill} handles the prompt and emits the first output token, and \emph{decode} then produces the remaining tokens, one in each step. Tokens already processed keep their attention keys and values in the key--value (KV) cache for reuse, so the cache footprint grows as generation advances~\cite{vidur24,vllm-sosp23}.

One GPU \emph{iteration} of a serving system executes work from several requests. With {\it continuous batching}, the batch can gain or lose requests between iterations~\cite{orca-osdi22}, and chunked prefill allows one iteration to mix prompt chunks with one token per decode request~\cite{sarathi-osdi24}. TTFT therefore covers both the wait and the prompt processing, and TPOT measures how fast later tokens are generated. The GPU's arithmetic throughput and memory bandwidth bound the iteration duration. Long prefills expose parallel computation, while small decode batches are held back by the reads of model weights and cached contexts~\cite{sarathi-osdi24,vllm-sosp23}. GPU memory also caps the KV cache, and with it how many requests can stay resident. When contexts no longer fit, the scheduler delays admission or preempts requests, whose cached context is discarded and recomputed later~\cite{vllm-sosp23}.

\noindent\textbf{Performance-modeling concepts.} ServeCast is specified as a \emph{queuing system}, a fundamental performance model that rests on the theory of Markov chains~\cite{bolch2006}. Its two main components are the \emph{server}, which sets how requests are scheduled and executed, and the \emph{workload}, which sets how external requests interact with the system.

A \emph{closed} workload has a fixed population of $N$ clients. Each client sends requests (i.e., prompts) to the server, waits for the response (i.e., the full output), and lets a \emph{think time} pass before the next request~\cite{lazowska-qsp84}. At most $N$ requests circulate in the system, so a larger $N$ at a fixed think time means a higher system load. Requests then compete increasingly for batch slots and KV-cache space, which reproduces the contention regimes seen in LLM serving.

Each request enters the model through its service demand, which here corresponds to its prompt and output lengths, the quantities that act directly on the prefill and decode stages. Classic queuing models assume exponential distributions, whereas we adopt the more expressive phase-type distributions~\cite{bolch2006}. Such a distribution expresses a random event, for instance the number of output tokens in a response, through explicit steps (i.e., the stages) visited according to a given probability distribution until a final state is reached. This well-known approach brings non-exponential distributions into a Markovian queuing framework, and ServeCast will also apply it to the non-exponential durations of batch iterations.

\section{The ServeCast Model}
\label{sec:model}
The setting is a single GPU on which vLLM runs with chunked prefill enabled~\cite{vllm-chunked-prefill-docs}. The serving configuration bounds the tokens and requests an iteration may select, as well as the available KV-cache capacity. The model includes eviction and recovery by recomputation~\cite{vllm-sosp23}.

Figure~\ref{fig:overview} gives an overview of ServeCast, with the modeled serving behavior on its left and the profiling and use of the model on its right. Figure~\ref{fig:overview}a draws the request lifecycle through the states the model tracks. A request leaves its thinking client and queues until its context fits in the KV cache of capacity $C$. It then executes in GPU iterations that combine prefill and decode work, until it completes and goes back to its client or, if memory runs out, is evicted and returns to the queue. Figure~\ref{fig:overview}b depicts the workflow that rests on this lifecycle. Offline, synthetic iterations of a source LLM, whether prefill, decode, or mixed, are profiled on the GPU (step~1), and the three timing coefficients are fitted to the measured durations (step~2). In the online step~3, the queueing model executes the lifecycle for a target LLM. Each iteration is timed by reusing the three coefficients fitted offline, combined with counts taken from the architecture of the target LLM and with GPU characteristics, so a new target requires no profiling of its own. With a trace-fitted workload as input, the model predicts the throughput, TTFT, and TPOT. The target may be the source LLM itself (RQ1) or a different LLM on the same GPU (RQ2). Sections~\ref{sec:workload-model}--\ref{sec:iteration-duration} detail each part, and the Appendix holds the formal model.

\subsection{Workload and Request Lengths}
\label{sec:workload-model}
As Section~\ref{sec:bg} anticipated, the workload is closed, with a population of $N$ clients. Think times are assumed exponential with mean $Z>0$. We use a state component in the model, denoted by $n_{\rm think}$,  which counts the thinking clients; so $N - n_{\rm think}$ is the number of clients wait for service. Further state components, discussed later, record the lifecycle stage of each request.

\begin{figure}[t]\centering
\begin{tikzpicture}[font=\scriptsize,
 phase/.style={circle,draw,minimum size=6.5mm,inner sep=0pt},
 endpoint/.style={draw,rounded corners,align=center,inner sep=3pt},
 arr/.style={-{Latex[length=1.3mm]},semithick},
 lab/.style={fill=white,inner sep=1pt}]
\draw[rounded corners] (1.7,-1.4) rectangle (5.85,2.05);
\node at (3.775,1.8) {Shared generation phases: $\Omega$};
\node[font=\scriptsize\itshape] at (4.45,1.2) {Branch 1};
\node[font=\scriptsize\itshape] at (4.45,-0.4) {Branch $b$};
\node at (0,1.8) {Prompt classes};
\node[phase] (entry) at (0,0.5) {$a$};
\node[phase] (entry2) at (0,-0.95) {$a'$};
\node at (0,-0.225) {$\vdots$};
\node[phase] (a1) at (2.2,0.5) {$1$};
\node[phase] (a2) at (3.5,0.5) {$2$};
\node (adots) at (4.35,0.5) {$\cdots$};
\node[phase] (a3) at (5.25,0.5) {$\ell$};
\node[phase] (b1) at (2.2,-0.95) {$r$};
\node (bdots) at (3.7,-0.95) {$\cdots$};
\node[phase] (b3) at (5.25,-0.95) {$L_g$};
\node at (3.7,-0.225) {$\vdots$};
\node[endpoint] (done) at (6.75,-0.225) {Complete};
\draw[arr,dashed] (entry) -- node[above,lab,pos=0.45] {$\pi_{a,1}$} (a1);
\draw[arr,dashed] (entry) -- (b1);
\draw[arr,dashed] (entry2) -- node[above,lab,pos=0.72] {$\pi_{a',1}$} (a1);
\draw[arr,dashed] (entry2) -- node[below,lab,pos=0.45] {$\pi_{a',r}$} (b1);
\draw[arr] (a1) -- node[below] {$\Omega_{12}$} (a2);
\draw[arr] (a2) -- (adots);
\draw[arr] (adots) -- (a3);
\draw[arr] (a3) -- node[above,lab] {$\omega_\ell$} (done);
\draw[arr] (b1) -- (bdots);
\draw[arr] (bdots) -- (b3);
\draw[arr] (b3) -- (done);
\draw[arr] (a1) edge[loop above] node {$\Omega_{11}$} (a1);
\foreach \n in {a2,a3,b1,b3}
  \draw[arr] (\n) edge[loop above] (\n);
\end{tikzpicture}
\caption{Prompt classes $a$, $a'$, $\ldots$ of the prompt/output model enter the shared generation branches with weights $\pi_{a,j}$ along the dashed arrows, and each solid transition emits one token.}
\label{fig:generation-phases}
\end{figure}
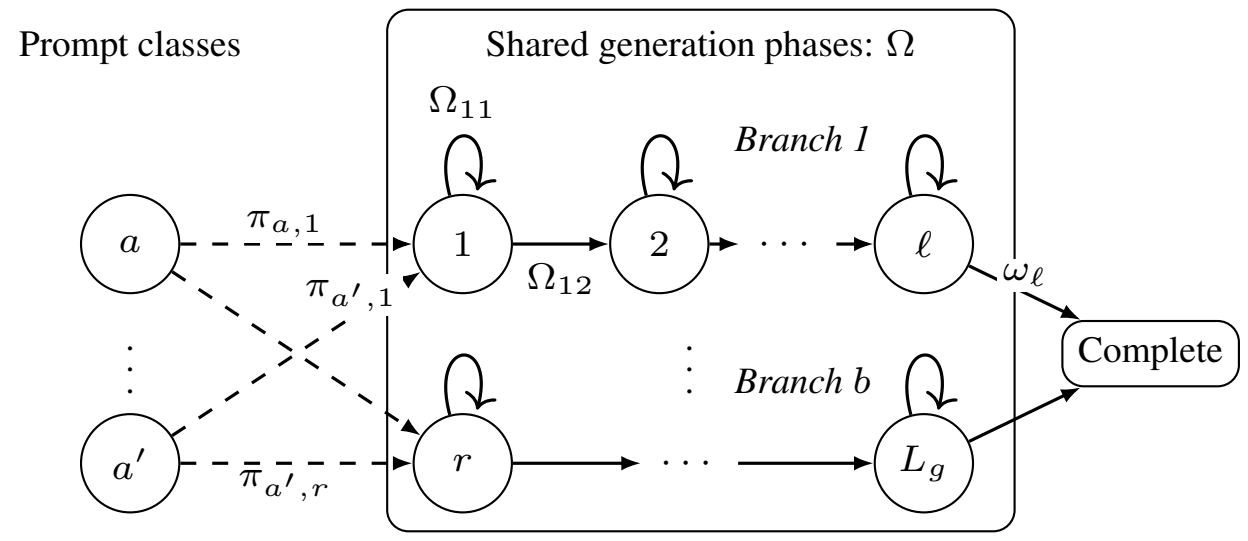

Prompt lengths are described by finitely many representative classes. Output length is modeled with a discrete phase-type distribution, since a request produces one output token per iteration. At every token the request either advances to the next generation phase or stays in its current one, and generation stops on leaving the last phase. Prompt/output correlations are carried by the entry probabilities. Here $\pi_{a,j}$ gives the probability that a request belongs to prompt class $a$ and, after its first output token, starts in generation phase $j$. The entry $j=0$ denotes a response that ends with that first token. For the other phases $j=1,\ldots,L_g$, the matrix $\Omega$ collects the probabilities of passing between generation phases. Its entry $\Omega_{jj'}$ is the probability of a move to phase $j'$ after one more token, and $\omega_j=1-\sum_{j'}\Omega_{jj'}$ is the probability that generation finishes at that token. The phases form a few parallel chains, called branches, and each request follows a single one of them, which makes the output length a mixture over branches (Figure~\ref{fig:generation-phases}). Every prompt class uses the same branches, and the classes are distinguished only by their entry probabilities. 

Retaining accuracy generally requires both detailed phase-type representations and the prompt/output dependence. The ablation tests of Section~\ref{sec:ablation} show that prediction quality can suffer when the dependence is dropped or when both lengths are set to their means~\cite{slim26,queueing-autoscaling26}.
\subsection{Model State and Evolution}
\label{sec:population-state}
\label{sec:transitions}
The state of ServeCast tracks requests along the lifecycle of Figure~\ref{fig:overview}a. Besides $n_{\rm think}$, it keeps waiting requests in the ordered queue $\mathcal Q$ and lists in $\mathcal P$ the prefill work scheduled for the coming iteration. It also holds the unfinished prefill $H_p$, and it groups the resident decode requests into the counts $n_{r,a,j,k}$ defined below. Each of the $n_p=|\mathcal P|$ entries stores its allocated tokens and cached progress, and $H_p$ carries an unfinished prefill request across iterations, together with its cached progress.

\paragraph{Request information and progress.}
A request is described by a record $u=(a,j,k)$ holding its prompt class $a$, its generation phase $j$ (Section~\ref{sec:workload-model}), and the number $k$ of tokens generated so far, with $k=0$ at submission. Its context length $p(u)=a+k$ determines the cache storage it needs. For an unfinished prefill, $H_p$ also records how many context tokens, $w\geq0$, have been processed and cached already.

The decode requests are grouped into the counts $n_{r,a,j,k}$. Their admission-age index $r\in\{0,\ldots,R_c-1\}$ is zero at admission and grows by one at each iteration until it reaches $R_c-1$, so the last group pools older admissions. Eviction picks its victims by this index. Summing the groups gives the decode request count $n_d$ and the cached decode context $K$, in tokens,
\begin{equation}
n_d=\sum_{r,a,j,k}n_{r,a,j,k},\qquad
K=\sum_{r,a,j,k}(a+k)n_{r,a,j,k}.\label{eq:decode-counts}
\end{equation}

\paragraph{Submissions and scheduling.}
Each submission lowers $n_{\rm think}$ by one and produces a request record $u=(a,j,0)$, with prompt length $a$ and generation phase $j$ sampled jointly with probability $\pi_{a,j}$. The request then waits in $\mathcal Q$ to be admitted.

Every iteration must respect the scheduler's token budget $B_{\rm tok}$ and the KV-cache capacity $C$. The latter, in token slots, depends on the memory given to the cache and on the storage each token of the LLM takes at its cache precision. As in the scheduling policy documented for vLLM~\cite{vllm-chunked-prefill-docs}, the model sets aside one token of $B_{\rm tok}$ for each resident decode request and assigns the remaining tokens to unfinished prefill first. The request in $H_p$, if it survives eviction, moves into $\mathcal P$ with its cached progress, its admission age, and its allocated prefill tokens.

The engine hands out KV-cache space in blocks of $B_{\rm blk}$ token slots, so the final block of a request may leave slots unused~\cite{vllm-sosp23}. For each resident decode request and each unfinished prefill request, ServeCast assumes $\sigma=(B_{\rm blk}-1)/2$ unused slots. Admission and eviction rely on the estimated occupancy, the sum of the cached context $K+w$ (where $w=0$ if $H_p$ is empty), the allowances above, and the growth the scheduled work requires.

When the coming decode tokens and the continuing prefill fit within $C$, the model scans $\mathcal Q$ from its head, subject to the remaining capacity and to the cap $B_{\rm seq}$ on the number of prefill and decode requests per iteration (vLLM's \texttt{max\_num\_seqs}). Admission compares the full context $p(u)$ of the candidate with the remaining capacity. An admitted candidate enters $\mathcal P$ with no cached progress and with its allocated prefill tokens. A failed check stops admission without giving the candidate any cache space, and the next admission attempt considers it again before the requests queued after it. Since prefill tokens are allocated in sequence, at most one request is left unfinished, held in $H_p$.

\paragraph{Eviction and reconstruction.}
When the coming decode tokens and the continuing prefill exceed $C$, the iteration admits no new request, and the model evicts requests until what remains fits. The first victim is the request in $H_p$, if any, which sequential prefill allocation makes the most recently admitted one, and $H_p$ is then cleared. If that is not enough, the model picks a decode request uniformly within the smallest occupied admission-age index $r$, whose count $n_{r,a,j,k}$ it decrements. In this way the age groups approximate reverse admission order, in line with the preference of vLLM for preempting requests that arrived later~\cite{vllm-sosp23}.

Whichever request is evicted gives back its estimated KV-cache occupancy and has its record $u=(a,j,k)$ placed at the head of $\mathcal Q$. Once readmitted, it enters $\mathcal P$ with $w=0$ and has to reconstruct all $p(u)=a+k$ of its context tokens.

\paragraph{Execution and completion.}
Completing an iteration changes $\mathcal P$, $H_p$, the decode counts $n_{r,a,j,k}$, and $n_{\rm think}$. An entry of $\mathcal P$ is written $(u_i,w_i,q_i,r_i)$, where $u_i=(a_i,j_i,k_i)$ is the request record, $w_i$ the cached tokens, $q_i$ the allocated prefill tokens, and $r_i$ the admission-age index.

Once processed, an entry leaves $\mathcal P$ with its cached progress raised to $w_i+q_i$. When $w_i+q_i<p(u_i)$, $H_p$ keeps $u_i$ with that progress and age $r_i$. Otherwise prefill is complete, and any matching entry in $H_p$ is cleared. After an initial prefill ($k_i=0$), the first output token ends the response when $j_i=0$, so $n_{\rm think}$ grows by one, and otherwise the request joins $n_{r_i,a_i,j_i,1}$. A completed reconstruction ($k_i>0$) instead increments $n_{r_i,a_i,j_i,k_i+1}$ and emits one more output token, with the generation phase $j_i$ unchanged.

A scheduled decode request is removed from its count $n_{r,a,j,k}$. With probability $\Omega_{jj'}$ it moves to $n_{r,a,j',k+1}$, and with probability $\omega_j$ it completes and $n_{\rm think}$ is incremented. The whole batch is updated jointly. After that, the admission-age indices of $H_p$ and of the resulting decode counts advance from $r$ to $\min(r+1,R_c-1)$.

\subsection{Iteration Duration}
\label{sec:iteration-duration}
Section~\ref{sec:transitions} gives the work an iteration selects and the state updates applied at its end, when every request selected for the batch advances. In its general formulation, the model represents the elapsed time by a shared Erlang clock, which through its stages can produce durations more concentrated than exponential ones~\cite{bolch2006,tijms2003}. The clock has mean $T$, computed from the prefill entries selected in $\mathcal P$, from the decode request count $n_d$ and the cached context $K$ given by Equation~\ref{eq:decode-counts}, and from the LLM and GPU parameters.

Specifically, the execution time $T$ has three explicit components, the dense layers, the vocabulary projection, and attention, to which overheads are added as in previous works~\cite{distserve-osdi24,imai2024}. A roofline-style estimate~\cite{williams-roofline09} of each component takes the larger of its modeled arithmetic time and its modeled memory-transfer time, as an approximation of their overlap~\cite{nvidia-gpu-performance}.

Let the GPU have memory bandwidth $B$, in bytes/s, and theoretical peak computation rate $F$, in floating-point operations/s, at the configured operating clock and arithmetic precision. The rate $F$ is scaled in proportion to the operating core frequency $f$. As in efficiency-based execution-time models~\cite{life25}, the effective computation rate of ServeCast is $\eta_c F$, with $\eta_c$ a dimensionless efficiency factor fitted to profiling measurements.

\paragraph{Scheduled tokens and dense layers.}
The prefill entries selected in $\mathcal P$ supply $q=\sum_{i=1}^{n_p}q_i$ tokens. Adding one token per decode request gives $q+n_d$ token positions for the linear layers of the decoder. The dense-layer estimate works with an effective token count $M$, which rounds $q+n_d$ up to the next configured CUDA-graph batch size, with $M=q+n_d$ beyond the largest such size.

Let $P$ be the number of parameters in the attention projections and the feed-forward transformations, and let $W$ be the bytes they occupy. The dense-layer estimate charges for reading these weights and for $2PM$ arithmetic operations, where a multiply-add counts as two. It adds to $T$ the term
\begin{equation}\label{eq:t1}
T_1=\max\left(\frac{W}{B},\frac{2PM}{\eta_cF}\right).
\end{equation}

\paragraph{Vocabulary projection.}
In the final projection, a hidden representation of width $d$ is mapped to scores over $V$ vocabulary tokens. To time it, ServeCast gives every selected request one row, for $R_h=n_p+n_d$ rows and $2dVR_h$ arithmetic operations. This approximation also counts unfinished prefill chunks, even though they emit no token. If $H$ is the number of bytes taken by the projection weights, the component contributes
\begin{equation}\label{eq:t2}
T_2=\max\left(\frac{H}{B},\frac{2dVR_h}{\eta_cF}\right).
\end{equation}
\paragraph{Attention.}
Attention time is estimated from prefill arithmetic and from reads of cached decode context. In prefill entry $i$, the $q_i$ new tokens each attend to the $w_i$ cached tokens and to the chunk tokens up to their own position. With the average context length approximated by $w_i+q_i/2$, the aggregate prefill work is $A=\sum_{i=1}^{n_p}q_i(w_i+q_i/2)$, and the estimated operation count over $\ell$ layers is $4d\ell A$.

Over all layers, one context token takes $C_{\rm KV}=2b_{\rm KV}\ell d_{\rm KV}$ bytes, with $b_{\rm KV}$ the element size and $d_{\rm KV}$ the aggregate width of the KV heads. In the bandwidth model, a single read of the cached decode context then costs $C_{\rm KV}K/B$ seconds.

Grouped-query attention lets $h_q$ query heads share $h_{\rm KV}$ key/value heads~\cite{ainslie-gqa23}, which gives $G=h_q/h_{\rm KV}$ query heads per KV head. If these query heads are processed separately, their shared context may have to be read repeatedly. ServeCast approximates reduced reuse under mixed prefill/decode execution with an effective multiplier on the KV-read cost,
\[
\gamma=\begin{cases}G,&q>0\text{ and }n_d>0,\\1,&\text{otherwise},\end{cases}\quad G=h_q/h_{\rm KV}.
\]

In pure decode, $\gamma=1$ assumes a single effective read of the cached context. In pure prefill $K=0$, and the decode-read term disappears. Attention therefore contributes to $T$ the term
\begin{equation}\label{eq:t3}
T_3=\max\left(\frac{\gamma C_{\rm KV}K}{B},\frac{4d\ell A}{\eta_cF}\right).
\end{equation}

\paragraph{Residual correction and total duration.}
The main arithmetic and memory costs are approximated by the three execution-time components. The remaining discrepancies are handled by a residual correction, affine in the number $n_d$ of decode requests, with coefficients $c_0$ and $c_d$. Both coefficients are in clock-normalized units, and the correction is assumed to scale with the inverse of the operating GPU core frequency $f$. In Section~\ref{sec:ablation} we test what each correction term contributes and whether prediction accuracy is preserved under the clock adjustment without a refit. The mean iteration duration is therefore
\begin{equation}
T=T_1+T_2+T_3+\frac{c_0+c_dn_d}{f}.\label{eq:iteration-cost}
\end{equation}
Execution-time profiling is needed only for the compute efficiency $\eta_c$ and for the residual coefficients $c_0$ and $c_d$, which are fitted as Section~\ref{sec:calibration} describes.

\section{Evaluation}
\label{sec:evaluation}
We answer RQ1--RQ3 from Section~\ref{sec:intro} by checking the TTFT, TPOT, and mean request throughput predicted by ServeCast against measured vLLM performance, our ground truth.
The baseline is Vidur~\cite{vidur24}, a state-of-the-art simulator in wide use, which predicts serving throughput and latency for a chosen model, GPU, and workload.

\subsection{Experimental Setup}
\label{sec:design}
Table~\ref{tab:setup} gives the models, GPUs, and traces, together with the serving settings. We run vLLM~0.25.1 with FlashAttention-2, which runs on all three GPUs. All model, GPU, and trace combinations are evaluated, with the sequence-limit exception noted in Table~\ref{tab:setup}.

On each model--GPU combination, ServeCast is evaluated under a closed workload whose think times are exponential with mean $Z=1$\,s. The client population $N$ takes six distinct levels, which span roughly 0.3--1.6 times our estimate of how many requests fit in the KV cache. Joint prompt/output lengths are fitted, as the next paragraph explains, to BurstGPT~\cite{burstgpt-kdd25}, to the Azure conversational trace published alongside Splitwise~\cite{splitwise-isca24}, and to ServeGen~\cite{servegen-nsdi26,acdc-sosp26}. This yields 324 \emph{configurations} in total, each given by a model, a GPU, a trace, and a client population $N$.

\begin{table}[t]\centering\small
\caption{Evaluated models, GPUs, and traces, with the serving settings.}\label{tab:setup}
\begin{tabular}{@{}p{.25\columnwidth}p{.68\columnwidth}@{}}\toprule
Factor & Levels or setting\\\midrule
Models & Llama-3.2-3B, Llama-3-8B; Qwen2.5-3B, Qwen2.5-7B; DeepSeek-Coder-1.3B, DeepSeek-LLM-7B\\
GPUs & A100 SXM4 40GB; H100 SXM5 80GB; RTX 3090 Ti 24GB\\
Traces & BurstGPT; Azure conversational; ServeGen\\
Serving & bfloat16; context limit 4096; token budget 2048; sequence limit 512 (256 for A100/Llama-3-8B/BurstGPT); KV block 16 tokens\\\bottomrule
\end{tabular}\end{table}

\paragraph{Prompt/output length estimation.}
\label{sec:calibration}
The representative prompt lengths and the probabilities of Section~\ref{sec:workload-model}, $\pi$ for entry and $\Omega$ for transitions, are estimated from all prompt/output pairs of each trace. Lloyd's algorithm~\cite{lloyd1982} condenses the observed prompt lengths into the fewest representative values $a$ that remain within 10\% of the observed distribution, in Wasserstein distance normalized by the mean prompt length. The discrete generation mixtures are fitted by expectation maximization, which gives $\pi$ and $\Omega$ (Appendix~C). Cross-validating on held-out pairs of prompt and output lengths selects the model structure, with $L_g\leq12$ and with fewer phases preferred at comparable predictive fit~\cite{hastie-esl09}.
Table~\ref{tab:workload-fit} summarizes the resulting workload structures, and Appendix~C details them.

\begin{table}[t]\centering\small
\caption{Structure of the fitted workloads.}\label{tab:workload-fit}
\begin{tabular}{@{}lrrr@{}}\toprule
 & BurstGPT & Azure & ServeGen\\\midrule
Prompt classes & 12 & 6 & 12\\
Branches & 4 & 2 & 2\\
Phases per branch & 1, 3, 5, 1 & 3, 8 & 1, 5\\
Total phases $L_g$ & 10 & 11 & 6\\
\bottomrule
\end{tabular}

\end{table}

\paragraph{Timing estimation.} The coefficients $(\eta_c,c_0,c_d)$ in Equation~\ref{eq:iteration-cost} are fitted, per model--GPU pair, to execution times measured on synthetic prefill, decode, or mixed batches. The probes vary the prefill chunk size, the number of decode requests, and the cached-context length. After warm-up, measurements are repeated with the GPU core clock fixed, and the coefficients come from weighted nonlinear least squares on relative errors in duration. Section~\ref{sec:rq3} reports, per GPU, the median and range of training MAPE over its six model-specific fits along with the profiling and simulation costs. Appendix~B gives the details.

\paragraph{Numerical prediction and measurement.}
Serving measurements and both simulators use the same transient-removal and precision routines. After the initial transient is discarded, nonoverlapping batch means~\cite{glynn-iglehart90} give 95\% confidence intervals, and each run goes on until the half-width of every interval is at most 5\% of the mean for TPOT and inverse throughput and at most 10\% for TTFT.
Measurements are taken on the GPUs listed in Table~\ref{tab:setup}, and one Apple M1 Pro laptop computes all ServeCast and Vidur predictions.

\subsection{RQ1: Prediction Accuracy}
\label{sec:rq1}
ServeCast and Vidur are both profiled on each target model--GPU pair, with workloads and KV capacity matched to the measurements. For every configuration and metric (request throughput, TTFT, TPOT) we take the absolute percentage error, and Table~\ref{tab:rq1} averages these errors by GPU, by model, and by workload.

\begin{table}[t]\centering\small
\caption{ServeCast and Vidur MAPE (\%), each profiled on the model--GPU pair it predicts. Shading and bold mark the best values.}\label{tab:rq1}
\setlength{\tabcolsep}{3.5pt}
\begin{tabular}{@{}cl rrr@{\hspace{9pt}}rrr@{}}\toprule
 & & \multicolumn{3}{c}{ServeCast} & \multicolumn{3}{c}{Vidur}\\\cmidrule(lr){3-5}\cmidrule(l){6-8}
 & & Thr. & TTFT & TPOT & Thr. & TTFT & TPOT\\\midrule
 & \textbf{All} & \winS{6.4} & \winS{11.9} & \winS{7.5} & 8.9 & 25.0 & 8.6 \\
\midrule
\multirow{3}{*}{\rotatebox[origin=c]{90}{\scriptsize GPU}} & A100 & \winS{4.0} & \winS{9.3} & \winS{4.9} & 10.7 & 27.4 & 9.8 \\
 & H100 SXM & \winS{5.9} & \winS{12.3} & \winS{7.9} & 7.5 & 26.0 & 8.3 \\
 & RTX 3090 Ti & 9.2 & \winS{14.0} & 9.5 & \winV{8.6} & 21.6 & \winV{7.7} \\
\midrule
\multirow{6}{*}{\rotatebox[origin=c]{90}{\scriptsize Model}} & Llama-3.2-3B & \winS{8.1} & \winS{10.3} & 9.4 & 8.5 & 26.6 & \winV{8.1} \\
 & Llama-3-8B & \winS{6.0} & \winS{11.4} & 7.4 & 7.5 & 22.9 & \winV{6.7} \\
 & Qwen2.5-3B & \winS{7.4} & \winS{10.3} & \winS{7.9} & 17.4 & 32.5 & 15.9 \\
 & Qwen2.5-7B & \winS{5.8} & \winS{11.4} & \winS{6.6} & 10.4 & 24.3 & 9.1 \\
 & DS-Coder-1.3B & 7.4 & \winS{13.4} & 9.0 & \winV{5.0} & 23.5 & \winV{6.8} \\
 & DS-LLM-7B & \winS{3.6} & \winS{14.5} & \winS{4.5} & 4.8 & 20.3 & 5.1 \\
\midrule
\multirow{3}{*}{\rotatebox[origin=c]{90}{\scriptsize Workload}} & BurstGPT & \winS{4.2} & \winS{8.6} & \winS{5.2} & 10.2 & 19.1 & 9.5 \\
 & Azure conv. & \winS{3.8} & \winS{6.8} & \winS{4.7} & 8.9 & 18.5 & 8.7 \\
 & ServeGen & 11.1 & \winS{20.3} & 12.5 & \winV{7.7} & 37.5 & \winV{7.6} \\
\bottomrule

\end{tabular}\end{table}

\rqanswer{Answer to RQ1}{Profiled on the target, ServeCast has a MAPE of 6.4\%, 11.9\%, and 7.5\% for throughput, TTFT, and TPOT, respectively, against 8.9\%, 25.0\%, and 8.6\% for Vidur. On every GPU and model, ServeCast has the lower TTFT error. Vidur has lower errors for throughput and TPOT on ServeGen, and also on the RTX 3090 Ti by smaller margins.}

\begin{figure*}[!t]\centering
\includegraphics[width=\textwidth]{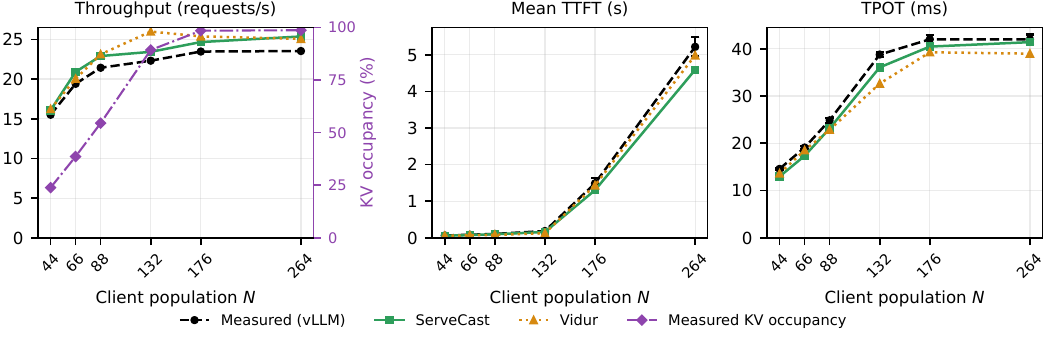}
\par\vspace{-6pt}
\caption{Example with Llama-3-8B, the H100 SXM, and BurstGPT, representative of the predominant directions of prediction error. The purple curve (right axis) is the mean usage of allocated KV capacity during the measurement windows, as sampled by the scheduler.}\label{fig:accuracy}\end{figure*}

Figure~\ref{fig:accuracy} uses Llama-3-8B with BurstGPT on the H100 SXM to show how the client population $N$ shapes system behavior and prediction accuracy. A larger $N$ lifts the measured KV-cache occupancy (the share of KV-cache capacity held by running requests) from 24\% to 99\%, as throughput nears a plateau and TTFT climbs steeply. This behavior is representative of every model--GPU--workload combination, across which the median KV-cache occupancy goes from 25.1\% (range 14.9--31.2\%) at the smallest $N$ to 98.9\% (93.7--99.7\%) at the largest. In this example ServeCast has the lower MAPE on all three metrics, although Vidur's TTFT prediction is more accurate at the largest $N$.

At all six populations, both predictors overestimate throughput and underestimate TTFT and TPOT, in line with the error directions that prevail across the evaluation. Over all tested configurations, ServeCast overestimates throughput in 82\% of cases and underestimates TTFT and TPOT in 62\% and 88\%, respectively, while the corresponding shares for Vidur are 77\%, 87\%, and 81\%.

\subsection{RQ2: \texorpdfstring{Cross-Model Transfer}{Cross-Model Transfer}}
\label{sec:rq2}
Cross-model transfer freezes the timing parameters $(\eta_c, c_0, c_d)$ fitted on one GPU--model pair (Section~\ref{sec:calibration}) and updates all other architectural parameters in Equations (\ref{eq:t1})--(\ref{eq:t3}), with no profiling on the target model. Vidur is not designed to carry profiles over from one model to another, so the reference remains target-profiled Vidur, as in RQ1. Table~\ref{tab:rq3-matrix} shows TTFT, whose aggregate error is the largest of the three metrics, and the appendix holds the matrices for throughput and TPOT.

\begin{table*}[t]\centering\small
\caption{ServeCast TTFT MAPE (\%) when transferred between models on the same GPU. Values are averaged first over each trace's six client populations and then across the three traces. Rows give the source model that was profiled and fitted, grouped by GPU, and each column is the target model. The light blue diagonal is the target-profiled reference of RQ1. In each GPU group, the last two rows set ServeCast, averaged over the five other source models, against target-profiled Vidur, and green (ServeCast) or amber (Vidur) shows which error is lower. In the final column, source rows average over the five off-diagonal targets and comparison rows over all six targets.}\label{tab:rq3-matrix}
\setlength{\tabcolsep}{3.5pt}
\begin{tabular}{@{}llcccccc@{\hspace{8pt}}c}\toprule
 & & \multicolumn{6}{c}{Predicted for} & \\\cmidrule(lr){3-8}
GPU & Profiled on & Llama-3.2-3B & Llama-3-8B & Qwen2.5-3B & Qwen2.5-7B & DS-Coder-1.3B & DS-LLM-7B & \begin{tabular}{@{}c@{}}Mean over\\models\end{tabular}\\\midrule
\multirow{8}{*}{\rotatebox[origin=c]{90}{A100}} & Llama-3.2-3B & \arxivdiag{6.9} & 11.8 & 8.5 & 12.4 & 13.9 & 11.6 & 11.6 \\
 & Llama-3-8B & 6.4 & \arxivdiag{9.9} & 7.3 & 10.4 & 9.4 & 11.4 & 9.0 \\
 & Qwen2.5-3B & 8.4 & 15.9 & \arxivdiag{7.5} & 18.2 & 11.8 & 13.2 & 13.5 \\
 & Qwen2.5-7B & 7.9 & 10.3 & 9.1 & \arxivdiag{10.5} & 13.1 & 10.3 & 10.1 \\
 & DS-Coder-1.3B & 16.5 & 31.3 & 14.8 & 34.7 & \arxivdiag{10.3} & 20.9 & 23.6 \\
 & DS-LLM-7B & 5.2 & 12.7 & 6.5 & 13.9 & 12.9 & \arxivdiag{10.9} & 10.2 \\
\cmidrule(l){2-9}
 & ServeCast, transferred & \winS{8.9} & \winS{16.4} & \winS{9.2} & \winS{17.9} & \winS{12.2} & \winS{13.5} & \winS{13.0} \\
 & Vidur, profiled & 28.5 & 27.4 & 33.9 & 27.6 & 25.2 & 22.0 & 27.4 \\
\midrule
\multirow{8}{*}{\rotatebox[origin=c]{90}{H100 SXM}} & Llama-3.2-3B & \arxivdiag{13.6} & 9.9 & 17.3 & 10.5 & 18.7 & 13.6 & 14.0 \\
 & Llama-3-8B & 13.3 & \arxivdiag{10.1} & 16.4 & 9.5 & 13.6 & 14.6 & 13.5 \\
 & Qwen2.5-3B & 9.4 & 11.7 & \arxivdiag{10.6} & 14.7 & 13.7 & 12.3 & 12.4 \\
 & Qwen2.5-7B & 15.6 & 11.0 & 19.3 & \arxivdiag{11.1} & 17.3 & 14.9 & 15.6 \\
 & DS-Coder-1.3B & 11.7 & 26.6 & 11.8 & 32.0 & \arxivdiag{15.0} & 16.4 & 19.7 \\
 & DS-LLM-7B & 12.0 & 10.4 & 14.8 & 9.4 & 10.6 & \arxivdiag{13.7} & 11.4 \\
\cmidrule(l){2-9}
 & ServeCast, transferred & \winS{12.4} & \winS{13.9} & \winS{15.9} & \winS{15.2} & \winS{14.8} & \winS{14.4} & \winS{14.4} \\
 & Vidur, profiled & 26.3 & 23.0 & 35.3 & 26.4 & 24.8 & 20.3 & 26.0 \\
\midrule
\multirow{8}{*}{\rotatebox[origin=c]{90}{RTX 3090 Ti}} & Llama-3.2-3B & \arxivdiag{10.3} & 18.2 & 13.2 & 22.5 & 18.7 & 21.8 & 18.9 \\
 & Llama-3-8B & 10.7 & \arxivdiag{14.3} & 11.5 & 16.6 & 16.5 & 19.2 & 14.9 \\
 & Qwen2.5-3B & 11.6 & 19.6 & \arxivdiag{12.9} & 22.7 & 16.9 & 18.8 & 17.9 \\
 & Qwen2.5-7B & 15.0 & 10.8 & 16.2 & \arxivdiag{12.7} & 23.0 & 17.6 & 16.5 \\
 & DS-Coder-1.3B & 14.2 & 22.1 & 15.7 & 27.5 & \arxivdiag{14.9} & 21.5 & 20.2 \\
 & DS-LLM-7B & 10.9 & 16.7 & 12.6 & 21.9 & 13.0 & \arxivdiag{18.9} & 15.0 \\
\cmidrule(l){2-9}
 & ServeCast, transferred & \winS{12.5} & \winS{17.5} & \winS{13.9} & 22.2 & \winS{17.6} & 19.8 & \winS{17.2} \\
 & Vidur, profiled & 25.1 & 18.2 & 28.4 & \winV{18.8} & 20.4 & \winV{18.6} & 21.6 \\
\bottomrule

\end{tabular}\end{table*}

\rqanswer{Answer to RQ2}{With cross-model transfer, the MAPE of ServeCast is 7.7\% for throughput, 14.9\% for TTFT, and 8.7\% for TPOT, against 8.9\%, 25.0\%, and 8.6\% for target-profiled Vidur.}

Transfer, as expected, usually costs some accuracy. The TTFT error of transferred ServeCast exceeds that of target-profiled ServeCast on every GPU (Table~\ref{tab:rq1}). Where another model's coefficients beat the target's own, the improvement is at most 4.4 points (DS-Coder-1.3B on the H100 SXM predicted from DS-LLM-7B, 10.6\% versus 15.0\%). That lies inside the 10\% half-width of the TTFT estimates, so we attribute it to noise. The hardest GPU is the RTX 3090 Ti, whose errors are the highest with transfer and without it. Of the source models, DS-Coder-1.3B transfers worst, and its mean error is the highest on every GPU, but it is no harder to predict from the other models than the rest are.

\begin{table}[t]\centering\small
\caption{Cost of profiling and prediction on each GPU. Training MAPE is the median and range of the iteration-duration error over the six model-specific fits of ServeCast, and profiling is the median GPU-minutes per model. Simulation averages the wall-clock time per configuration over the 108 configurations of the GPU, with both simulators on one machine, and the speedup divides the mean time of Vidur by that of ServeCast. Shading marks the lower simulation time. S is ServeCast and V is Vidur.}\label{tab:rq2-cost}
\setlength{\tabcolsep}{1.5pt}
\begin{tabular}{@{}lrr@{\hspace{7pt}}rr@{\hspace{7pt}}rrr@{}}\toprule
 & \multicolumn{2}{c}{\shortstack{Training\\MAPE (\%)}} & \multicolumn{2}{c}{\shortstack{Profiling\\(GPU-min)}} & \multicolumn{3}{c}{Simulation (s)}\\
\cmidrule(r{5pt}){2-3}\cmidrule(r{5pt}){4-5}\cmidrule{6-8}
GPU & Median & Range & S & V & S & V & Speedup\\\midrule
A100 & 9.4 & 8.0--11.4 & 3.1 & 18.1 & \winS{12.0} & 29.0 & 2.4$\times$ \\
H100 SXM & 5.0 & 4.7--9.0 & 2.0 & 13.3 & \winS{11.1} & 27.5 & 2.5$\times$ \\
RTX 3090 Ti & 10.7 & 8.6--13.0 & 4.6 & 13.1 & \winS{16.1} & 21.2 & 1.3$\times$ \\
\bottomrule

\end{tabular}\end{table}

\subsection{RQ3: \texorpdfstring{Profiling}{Profiling} and Prediction Cost}
\label{sec:rq3}
For each GPU, Table~\ref{tab:rq2-cost} lists the median time to profile a model and the mean time to simulate a configuration. The measured profiling costs of ServeCast are lower. The two methods, however, profile with different settings, each fixed for the whole evaluation, so under matched conditions these costs establish no profiling speedup.

\rqanswer{Answer to RQ3}{On average, ServeCast takes 13.1\,s of simulation time and Vidur 25.9\,s, a 2.0$\times$ speedup, and ServeCast is faster on every GPU.}

\subsection{Ablation}
\label{sec:ablation}
Three components of ServeCast are ablated, namely the distribution of prompt/output lengths, the residual correction of the timing model in Equation~\ref{eq:iteration-cost}, and how that model depends on the core clock $f$.

\paragraph{Prompt/output length distribution.} The request-length model of Section~\ref{sec:workload-model} is simplified in two ways. \emph{Independent lengths} substitutes $(\sum_{j'}\pi_{a,j'})(\sum_{a'}\pi_{a',j})$ for $\pi_{a,j}$ and leaves $\Omega$ untouched, so prompt and output lengths keep their fitted distributions but are no longer dependent. \emph{Fixed mean lengths} gives each request lengths equal to the fitted means of $a$ and $g$, rounded, which removes length variability too.

\begin{table}[t]\centering\small
\caption{Ablation of prompt/output lengths. MAPE (\%) over the six client populations of each model--GPU--workload combination, sorted by the TTFT error of the original model.}\label{tab:ablation}
\begin{tabular}{@{}lrrr@{}}\toprule
Length model & Thr. & TTFT & TPOT\\\midrule
\multicolumn{4}{@{}l@{}}{\textit{A100 / Llama-3.2-3B / Azure}}\\
Original joint distribution & 2.9 & 0.9 & 3.6\\
Independent lengths & 5.2 & 15.3 & 3.8\\
Fixed mean lengths & 10.1 & 25.9 & 7.5\\
\midrule
\multicolumn{4}{@{}l@{}}{\textit{A100 / Qwen2.5-7B / Azure}}\\
Original joint distribution & 0.5 & 10.7 & 1.4\\
Independent lengths & 1.7 & 16.2 & 3.5\\
Fixed mean lengths & 6.9 & 93.9 & 8.6\\
\midrule
\multicolumn{4}{@{}l@{}}{\textit{RTX 3090 Ti / DS-Coder-1.3B / ServeGen}}\\
Original joint distribution & 20.7 & 28.7 & 20.6\\
Independent lengths & 21.3 & 29.2 & 19.8\\
Fixed mean lengths & 48.7 & 56.7 & 37.1\\
\bottomrule\end{tabular}

\end{table}

This ablation covers the model--GPU--trace combinations of RQ1 that have the minimum, upper-median, and maximum TTFT MAPE, and Table~\ref{tab:ablation} sets the two simplifications against the original joint distribution.
With lengths fixed at their means, all three errors grow substantially in every selected combination, and TTFT MAPE reaches 93.9\%. On the two Azure combinations, removing the dependence also raises every error. On ServeGen its effect is small, which agrees with the nearly independent fitted distribution in those runs.

\paragraph{Residual correction.} The constant and the per-decode-request contributions of Equation~\ref{eq:iteration-cost} are tested next. \emph{Constant only} keeps $c_0/f$ and sets $c_d=0$, while \emph{no correction} sets $c_0=c_d=0$. In each variant, the remaining timing coefficients are refitted with the same procedure (Section~\ref{sec:design}).

\begin{table}[t]\centering\small
\caption{Ablation of the residual correction, with MAPE (\%) averaged over model--GPU--workload combinations. Parentheses hold the share of evaluations in which the absolute percentage error exceeds that of the full model.}\label{tab:residual-ablation}
\setlength{\tabcolsep}{3pt}
\begin{tabular}{@{}lrrr@{}}\toprule
Residual correction & Thr. & TTFT & TPOT\\\midrule
\multicolumn{4}{@{}l@{}}{\textit{A100}}\\
Full & 4.0\phantom{ (00.0)} & 9.3\phantom{ (00.0)} & 4.9\phantom{ (00.0)}\\
Constant only & 4.7 (64.8) & 9.2 (50.5) & 5.6 (73.3)\\
No correction & 18.2 (89.5) & 16.0 (70.5) & 16.0 (86.7)\\
\midrule
\multicolumn{4}{@{}l@{}}{\textit{H100 SXM}}\\
Full & 5.9\phantom{ (00.0)} & 12.3\phantom{ (00.0)} & 7.9\phantom{ (00.0)}\\
Constant only & 7.3 (78.5) & 12.8 (61.7) & 9.3 (85.0)\\
No correction & 31.9 (90.7) & 19.5 (68.2) & 28.9 (94.4)\\
\midrule
\multicolumn{4}{@{}l@{}}{\textit{RTX 3090 Ti}}\\
Full & 9.2\phantom{ (00.0)} & 14.0\phantom{ (00.0)} & 9.5\phantom{ (00.0)}\\
Constant only & 11.2 (74.5) & 15.2 (67.9) & 10.8 (66.0)\\
No correction & 24.0 (84.0) & 21.1 (80.2) & 19.0 (81.1)\\
\midrule
\multicolumn{4}{@{}l@{}}{\textit{All GPUs}}\\
Full & 6.4\phantom{ (00.0)} & 11.9\phantom{ (00.0)} & 7.5\phantom{ (00.0)}\\
Constant only & 7.7 (72.6) & 12.4 (60.1) & 8.6 (74.8)\\
No correction & 24.7 (88.1) & 18.9 (73.0) & 21.3 (87.4)\\
\bottomrule\end{tabular}

\end{table}

Table~\ref{tab:residual-ablation} covers all 54 model--GPU--trace combinations of RQ1. Without the correction, all three MAPEs rise substantially on every GPU, and throughput MAPE reaches 31.9\% on the H100. Most of the accuracy comes back with the constant term alone. Adding the per-decode-request term then lowers what error remains on every GPU and every metric, except for TTFT on the A100, where the two variants are 0.1 points apart.

\paragraph{Clock dependence.} To save energy, serving systems reduce GPU frequencies~\cite{dynamollm-hpca25,beam26}. We therefore ask if ServeCast, through the explicit clock dependence in Eq.~(\ref{eq:iteration-cost}), can predict performance at another core clock with no refit of its timing coefficients. In the full model, the arithmetic rate $F$ grows in proportion to $f$, while the residual correction varies inversely with the clock frequency $f$. Two variants are compared with it. The first updates only $F$ and holds the residual at its value for the profiling clock, and the second changes neither. As a reference, a model is profiled and refitted at every target clock. With the 1200\,MHz parameters retained, we evaluate Llama-3-8B with Azure on the A100 at core clocks locked to 900 and 750\,MHz and an unchanged memory clock.

\begin{table}[t]\centering\small
\caption{Clock-dependence ablation for Llama-3-8B and Azure on the A100, with MAPE (\%) over the client populations $N$ at each core clock. The three clock-adjustment variants keep the coefficients fitted at 1200\,MHz, and the reference is refitted at every target clock.}\label{tab:clock-ablation}
\begin{tabular}{@{}lrrr@{}}\toprule
Clock treatment & Thr. & TTFT & TPOT\\\midrule
\multicolumn{4}{@{}l@{}}{\textit{900\,MHz}}\\
Adjust $F$ and residual & 2.3 & 10.3 & 2.2\\
Adjust $F$ only & 1.8 & 9.6 & 3.7\\
No clock adjustment & 19.5 & 14.7 & 17.9\\
Refit at target clock & 1.8 & 9.3 & 2.2\\
\midrule
\multicolumn{4}{@{}l@{}}{\textit{750\,MHz}}\\
Adjust $F$ and residual & 1.6 & 8.5 & 3.0\\
Adjust $F$ only & 6.0 & 8.5 & 7.1\\
No clock adjustment & 38.4 & 29.0 & 29.4\\
Refit at target clock & 2.0 & 7.2 & 2.9\\
\bottomrule\end{tabular}

\end{table}

Table~\ref{tab:clock-ablation} averages the errors over the two clocks. With both terms updated, TPOT MAPE is 2.6\%, compared with 5.4\% when only $F$ is updated, 23.6\% without adaptation, and 2.5\% with a refit at that clock. With both terms updated, the throughput and TTFT errors stay no more than 1.5 points from the refitted reference.

\subsection{Limitations}
\label{sec:limitations}
We acknowledge several limitations of this work.
Chief among them is the scope of the evaluation. The models we evaluate have at most 8B parameters, and each is served on one GPU with no tensor parallelism. Production deployments often run much larger models over several GPUs. Cross-model transfer should pay off most there, because profiling and measuring such models demands multi-GPU machines and much more time and money. The same cost confined our evaluation to the single-GPU scale reported here, and extending ServeCast to tensor-parallel serving is our first direction for future work. The timing coefficients are another limitation, since they are fitted and not measured and they differ from model to model (Appendix~B.4). Transfer bears the cost of this difference, most visibly when the source is DS-Coder-1.3B (Table~\ref{tab:rq3-matrix}).

\section{Related Work}
\label{sec:related}
\paragraph{Serving simulators.}
For its serving simulation, Vidur trains runtime predictors on operator profiles of each model on each GPU \cite{vidur24}. Its timings thus belong to the profiled model, and another model needs a new profile. In our evaluation, Vidur's TTFT error is roughly twice ServeCast's. KernelSight-LM adds to Vidur a roofline kernel model, learns its efficiency for each family with boosted trees or small neural networks, and transfers the model to an unseen GPU generation \cite{kernelsight-lm26}. By contrast, ServeCast fits three coefficients attached to counted architecture quantities and transfers them between models on the same GPU. The KernelSight-LM paper also reports that its TTFT error climbs from 15--22\% to 74\% once the offered load passes capacity, a regime it regards as rare. In a what-if analysis it is unknown beforehand whether a configuration saturates, so our evaluation covers client populations reaching 1.6 times the KV-cache capacity. For the two largest populations, the MAPE of ServeCast stays comparable to that at lighter loads (Appendix~E.3), at 5.6\%, 11.7\%, and 7.4\% for throughput, TTFT, and TPOT, respectively, so ServeCast is suited to this kind of analysis. Frontier extends simulation with detailed scheduling, memory, and distributed execution, and it shows that average-case proxies distort the predictions \cite{frontier-sim26}, as does our length ablation. Beyond a single GPU, APEX, Charon, and LLMServingSim explore parallel execution, large-scale training and inference, and the design of accelerators \cite{apex,charon26,llmservingsim}. ServeCast models one GPU for now, and carrying its counted-work cost rule over to multi-GPU deployments remains future work.

\paragraph{Analytical serving models.}
In SLIM and the model of Ramani and Tantawi, every request carries the mean input and output lengths \cite{slim26,queueing-autoscaling26}, while Ray et al. fit a throughput curve for each combination of input and output lengths \cite{ala25}. All three thus describe a workload through fixed lengths. This choice appears costly in our length ablation, where fixed mean lengths push the TTFT MAPE of ServeCast up to 93.9\%. ServeCast instead fits the prompt and output lengths with a joint distribution, and makes admission, batching, eviction, and recomputation depend on the progress of each request.

\paragraph{Execution-time models and \texorpdfstring{transfer across models}{transfer across models}.}
In LIFE and GenZ, every operator is priced by a roofline scaled with measured or assumed efficiencies \cite{life25,genz24}. NeuSight forecasts unseen GPUs by using performance laws to bound learned tile-level predictions, and LLM-Viewer and LLMCompass locate the bottlenecks of inference and guide the design of accelerators \cite{neusight-asplos25,llmviewer24,llmcompass-isca24}. These tools each price a given configuration, while serving performance depends on how the batch changes as requests are admitted, evicted, and completed. Offline, SOLA fits polynomials of per-iteration latency, and online it rescales them with measured latency for scheduling a running system \cite{sola25}. ServeCast predicts how the batch changes and prices every iteration by the roofline form given above plus a fitted residual that our ablation finds necessary. The search over serving configurations in AIConfigurator and the deployment estimator of Cho et al. draws on a calibrated kernel-level database or on more than 100K interpolated operator measurements per GPU \cite{aiconfigurator26,meta-deploy26}. For each GPU, ServeCast needs just its three fitted coefficients.

\section{Conclusion}
To predict LLM serving performance, ServeCast couples the progression of requests and memory constraints with an iteration-time model that keeps architectural inputs apart from fitted timing coefficients. This separation allows transfer between LLMs on the same GPU with no timing profiling of the target model. With ServeCast and Vidur both calibrated on the target model--GPU pair, ServeCast has the lower aggregate prediction errors, and its margin is largest for TTFT. Cross-model transfer raises the error on average, yet keeps aggregate TTFT error below that of target-calibrated Vidur, with comparable throughput and TPOT errors. Reusable calibration and lower average simulation times together make ServeCast a practical tool for capacity planning and for what-if analysis over candidate LLMs.

This work covers closed workloads on one GPU shared by prefill and decode work, and evaluates mean request throughput, TTFT, and TPOT. The queueing formulation is a basis for widening this scope. Future work could add open arrivals whose interarrival times are phase-type and take the state and latency analysis further to predict tail distributions for assessing SLOs. Queueing networks could also replace the single GPU to represent GPU clusters with separate servers for prefill and decode.

\label{maintextend}
\balance
\bibliography{servecast}

\section*{AI Usage Disclosure}
We used an AI coding assistant (Claude Code with Anthropic Claude models) to write and test parts of the measurement harness and the analysis code, and to edit the manuscript text for clarity and length. The authors designed the model and the experiments, checked every result and citation, and take full responsibility for the content.
\clearpage
\onecolumn
\arxivappendixfloats
\raggedbottom
\allowdisplaybreaks[1]
\newcommand{\m}[1]{{\bfseries comment:} \emph{#1} {\bfseries end comment}}
\renewcommand{\arraystretch}{1.15}
\setlength{\LTcapwidth}{\textwidth}
\appendix
Section~\ref{app:formal-model} gives the formal model and Section~\ref{app:timing} the timing calibration. Section~\ref{app:workload-construction} covers trace-based workload fitting, Section~\ref{app:experiment-config} the experimental traces and cache capacities, and Section~\ref{app:complete-results} supplementary results. Unless the main paper is named, numbering is that of this appendix.

\section{Formal Model}
\label{app:formal-model}
Below we formalize the request lifecycle and the iteration timing of Sections~3.1--3.3 of the main paper.

\paragraph{Shared clock.}
The iteration-cost equation in Section~3.3 of the main paper gives the selected batch a mean duration $T$, and the batch keeps its composition while it executes. Execution runs through every stage $c\in\{1,\ldots,k_{\rm erl}\}$ of the shared Erlang clock, for an integer $k_{\rm erl}\geq1$. Submissions may arrive in any stage, the last included. The iteration is complete once the clock exits the last stage. Each stage has rate $\lambda_T=k_{\rm erl}/T$, where the selected batch fixes $T$.

\subsection{Model State}
\label{app:state}
The client population, the request records $u=(a,j,k)$, and the generation law follow Sections~3.1--3.2 of the main paper. Submissions wait in two ordered staging lists $\mathcal S,\widehat{\mathcal S}$ until they become eligible for scheduling (Section~\ref{app:eligibility}). An eligible request sits in the waiting queue $\mathcal Q$ or in the blocked-candidate slot $B_q$, which holds a candidate the admission memory check has refused, and $B_q=\varnothing$ when the slot is empty. Together the four components form the logical waiting queue of the main paper.

Resident progress is kept as an unfinished prefill or in the decode counts. If an iteration ends before initial prefill or reconstruction is done, $H_p=(u,w,r)$ keeps the cached progress $1\leq w<p(u)$ and the admission-age index $r$, and $H_p=\varnothing$ once no prefill is left unfinished. Decode residents use the main paper's counts $n_{r,a,j,k}$, where $r\in\{0,\ldots,R_c-1\}$, $j\in\{1,\ldots,L_g\}$, and $k\geq1$. The totals $n_d=\sum_{r,a,j,k}n_{r,a,j,k}$ and $K$ for the decode population and the cached context are those of the main paper. The count vector is $\mathbf n$, with unit vectors $\mathbf e_{r,a,j,k}$.

The list $\mathcal P$ names the prefill tokens that move forward in the current iteration. Each entry $(u,w,q,r)$ holds the request, its cached token count, its allocated token count (the token grant), and its admission-age index, with $0\leq w<p(u)$ and $0<q\leq p(u)-w$. A continuing $H_p$ is in $\mathcal P$ only if its allocated token count $q$ is positive.

The state also includes the clock stage $c$, and an idle system has $c=0$ and an empty $\mathcal P$.

\paragraph{Complete state.}
Gathering the components gives the state
\begin{equation}
\mathcal X=(n_{\rm think},\mathcal Q,B_q,\mathcal S,\widehat{\mathcal S},H_p,\mathbf n,\mathcal P,c).
\end{equation}
Once the architecture, hardware, and configuration inputs are fixed, the selected batch yields $T=T(\mathcal P,\mathbf n)$. During execution, submissions alter only $n_{\rm think}$ and the waiting lists.

\subsection{Submission and Eligibility}
\label{app:eligibility}
Because the serving engine can prepare the next batch while the current one runs, a new submission becomes eligible only after a delay, which the staging lists approximate.

Under the main paper's assumption of exponential think times, submissions occur at total rate $n_{\rm think}/Z$. The rate in Table~\ref{tab:transitions} is this total times the request-type probability $\pi_{a,j}$.

A request submitted in iteration $i$ goes into $\mathcal S$. As long as resident work exists, meaning decode requests or an unfinished prefill, it passes to $\widehat{\mathcal S}$ after iteration $i$ and becomes eligible after iteration $i+1$, so iteration $i+2$ is the earliest that can select it. It still has to pass the admission checks before joining a batch.

With no resident work to continue, the delay is skipped. A submission in an idle system enters $\mathcal Q$ at once, and an iteration completion that leaves no resident work releases both staging lists.

Let $\|$ denote ordered concatenation and $[u]$ a one-request list. After the iteration's work updates and the advance of the admission-age indices (Section~\ref{app:replay}), the lists change as
\begin{equation}
(\mathcal Q,\widehat{\mathcal S},\mathcal S)\leftarrow
\begin{cases}
(\mathcal Q\|\widehat{\mathcal S},\mathcal S,\varnothing),&\text{resident work remains},\\
(\mathcal Q\|\widehat{\mathcal S}\|\mathcal S,\varnothing,\varnothing),&\text{otherwise}.
\end{cases}
\label{eq:staging-release}
\end{equation}
Neither list changes at an intermediate clock transition. At the scheduling boundary, where the scheduler picks the next batch, the release has already emptied $\mathcal S$.

\subsection{Scheduling, Admission, and Eviction}
\label{app:scheduling}
The token budget $B_{\rm tok}$ caps the tokens processed in one iteration. At every scheduling boundary ServeCast first sets aside one token per decode request and gives the tokens left over to a continuing $H_p$. If the cache cannot hold this progress, eviction frees room for the surviving residents to advance. Unfinished prefill is evicted first, then, as needed, decode requests from the youngest occupied admission-age group. When no eviction is required, whatever cache space and token budget remain may go to admitting waiting requests. No model time passes during these scheduling operations. Token allocation needs $n_d\leq B_{\rm tok}$, and every evaluated sequence limit satisfies $B_{\rm seq}\leq B_{\rm tok}$.

\paragraph{Resident memory check.}
Cache memory comes in blocks of $B_{\rm blk}$ token slots, so the last block of a request may hold unused slots. ServeCast approximates this unused space by a mean slack of $\sigma=(B_{\rm blk}-1)/2$ token slots per resident. We write $C$ for the usable cache capacity in token slots and $K$ for the decode residents' total cached context in tokens. The indicator $h$ is one when an unfinished prefill $H_p$ exists and zero otherwise, and that prefill has cached progress $w_H$ and required context length $p_H$. The current occupancy estimate $\mathrm{Occ}$ is
\begin{equation}
\mathrm{Occ}=K+n_d\sigma+h(w_H+\sigma).
\end{equation}
The grant $q_H$ of the continuing prefill cannot exceed its remaining work or the token budget that decode leaves, so
\begin{equation}
q_H=\min(p_H-w_H,B_{\rm tok}-n_d).
\end{equation}
If $H_p$ is absent, its occupancy term and $q_H$ vanish. The remaining processing budget is then $r_0$, with
\begin{equation}
r_0=B_{\rm tok}-n_d-q_H.
\end{equation}
The remaining cache space $F_{\rm rem}$ deducts the current occupancy and $n_d+q_H$ tokens of planned resident growth,
\begin{equation}
F_{\rm rem}=C-\mathrm{Occ}-n_d-q_H.
\end{equation}

\paragraph{Eviction order.}
If $F_{\rm rem}<0$, current occupancy and planned growth together exceed the usable cache capacity, and the scheduler evicts until the remaining residents pass the check. Each step evicts the unfinished prefill if there is one, and otherwise draws a request uniformly from the youngest occupied admission-age group, $r^\star=\min\{r:\sum_{a,j,k}n_{r,a,j,k}>0\}$. Since every request of that group is equally likely, a cell is chosen with probability proportional to its request count. Each eviction releases the victim's slack, context, and planned growth, after which grants and remaining capacity are recomputed.

Admission-age groups approximate admission order, and requests in one group are exchangeable. The configuration uses $R_c=n_{\rm recent}+2$ groups, one for current admissions, $n_{\rm recent}$ for past iterations kept apart, and one pooling the older ones. Every evaluated prediction takes $n_{\rm recent}=128$, hence $R_c=130$.

An evicted request rejoins the queue in front of any blocked candidate (Table~\ref{tab:transitions}). It keeps its record, and on readmission all $p(u)=a+k$ tokens must be reconstructed. An iteration with an eviction admits no new request, though surviving residents can still progress.

\paragraph{Admission.}
In an iteration without eviction, a request held in the blocked-candidate slot $B_q$ comes before the waiting queue $\mathcal Q$. Admission reserves memory for the full required context of a candidate even if the remaining token budget allows only a partial grant. With $x_i$ the required admission length of candidate $i$ and $P_i=\sum_{\nu=1}^ix_\nu$, the scheduler examines candidate $i$ only when $P_{i-1}<r_0$ and the configured sequence limit $B_{\rm seq}$ allows it. The admission memory check for that candidate reads
\begin{equation}P_i+i\sigma\leq F_{\rm rem}.\label{eq:admission-guard}
\end{equation}
Its token grant equals $\min(x_i,r_0-P_{i-1})$. A refusal by the memory check leaves a blocked candidate behind, and a partial grant terminates admission. With $r_0=0$, the scheduler examines no waiting candidate and creates no new blocked candidate. In particular, when $n_d=B_{\rm tok}$ the unfinished prefill gets zero tokens and, provided it passes the resident memory check, stays resident without progress.

\paragraph{Selected work.}
A surviving $H_p$ adds $(u,w,q_H,r)$ to $\mathcal P$ only if $q_H>0$, and a prefill without a token grant remains resident but has no timing row. New admissions follow Table~\ref{tab:transitions}. Granting tokens in sequence leaves at most one prefill unfinished. The main paper's iteration-cost rule then computes $T$ from the decode counts and the selected prefill work.

\subsection{Execution and Completion}
\label{app:replay}

Execution begins in clock stage $c=1$. In stage $c=k_{\rm erl}$, the jump rate of a joint work outcome $\xi$ is $\lambda_Tp(\xi\mid\mathcal X)$. The clock progression, the simultaneous work updates, the admission-age updates, the staging release, and the reset are given in Table~\ref{tab:transitions}.

\paragraph{Work outcomes.}
Every ordinary-decode request emits one token, then completes or enters its next phase. Take an occupied cell $s=(r,a,j,k)$ with count $n_s=n_{r,a,j,k}$ and let $C_{s,d}$ count its requests sent to destination $d$, where destination zero means completion and $1,\ldots,L_g$ are the generation phases. Independently for each cell,
\begin{equation}
(C_{s,0},\ldots,C_{s,L_g})\sim\operatorname{Multinomial}
(n_s;\omega_j,\Omega_{j1},\ldots,\Omega_{jL_g}).
\end{equation}
The resulting joint probability is $p(\xi\mid\mathcal X)$. Conditional on the counts, request identities within a cell are assigned to destinations uniformly without replacement. The solver equivalently draws destinations per phase total and then assigns identities uniformly without replacement, which keeps every request's context. Prefill and reconstruction updates follow deterministically from the selected work.

Initial prefill emits the first token on completion. A completing reconstruction emits one more token and keeps generation phase $j$ and the first token's timestamp (Table~\ref{tab:transitions}). Waiting after an eviction and the reconstruction itself therefore fall within the first-to-final-token span. An interrupted prefill or reconstruction emits nothing. In every reported prediction, reconstruction keeps the generation phase. Subjecting the emitted token to an ordinary phase transition would give a different completion model.

\paragraph{Occupancy after completion.}
The memory checks reserve resident growth together with each entrant's full required context and slack. A prefill or reconstruction that completes emits one token in excess of this reservation, and requests that complete free their contexts.

Let $n_{\rm adm}$ be the number of new admissions in $\mathcal P$, not counting the continuing $H_p$, and let $h$ indicate whether $H_p$ survives eviction. The number of completed prefills or reconstructions, the continuing prefill included, is $c_{\rm pf}$. After an iteration completes, the resident occupancy thus obeys
\begin{equation}
\mathrm{Occ}_{\rm end}\leq C+c_{\rm pf}\leq C+n_{\rm adm}+h.
\end{equation}
The inequality bounds the token-plus-mean-slack approximation with eviction enabled and gives no exact bound on block allocation. A completing unfinished prefill still contributes when the iteration admits no new request.

\paragraph{Transition specification.}
The updates are listed in Table~\ref{tab:transitions}. Boundary operations consume no model time. At iteration completion the work outcomes are applied together, and state components not listed keep their values.

\begingroup
\normalsize
\setlength{\tabcolsep}{4pt}
\begin{longtable}{@{}>{\raggedright\arraybackslash}p{.14\textwidth}>{\raggedright\arraybackslash}p{.54\textwidth}>{\raggedright\arraybackslash}p{.28\textwidth}@{}}
\caption{Timing and state updates. All work updates take place at once, before the admission-age updates and the staging-list release.}\label{tab:transitions}\\
\toprule
Event & State update & Rate or condition\\\midrule
\endfirsthead
\multicolumn{3}{l}{\tablename~\thetable\ continued}\\
\toprule Event & State update & Rate or condition\\\midrule
\endhead
\midrule\multicolumn{3}{r}{Continued on next page}\\
\endfoot
\bottomrule
\endlastfoot
Submission & $n_{\rm think}\leftarrow n_{\rm think}-1$; create waiting record $u=(a,j,0)$ under the eligibility rule. & Rate $n_{\rm think}\pi_{a,j}/Z$, $Z>0$.\\[3pt]
Admission & Remove candidate $u$ from $B_q$ or the front of $\mathcal Q$; append $(u,0,q,0)$ to $\mathcal P$. & Boundary; resident grants set, no eviction, $B_{\rm seq}$ and memory checks pass, $q>0$.\\[3pt]
Memory refusal & If $\mathcal Q=[u]\|\mathcal Q'$, set $\mathcal Q\leftarrow \mathcal Q'$, $B_q\leftarrow u$; an already blocked candidate stays in $B_q$. Stop admission. & Boundary; first failed memory check.\\[3pt]
Eviction & $H_p=(u,w,r)\to\varnothing$, or $\mathbf n\leftarrow\mathbf n-\mathbf e_{r^\star,a,j,k}$ with $u=(a,j,k)$. In either case, $\mathcal Q\leftarrow[u]\|[B_q]\|\mathcal Q$, $B_q\leftarrow\varnothing$; omit $[B_q]$ if empty. & Boundary; $F_{\rm rem}<0$. Selection and recomputation follow Section~\ref{app:scheduling}.\\[3pt]
Clock progression & $c\leftarrow c+1$, for $1\leq c<k_{\rm erl}$. & Rate $\lambda_T=k_{\rm erl}/T$.\\\midrule
Iteration completion & Apply all work updates below jointly, then perform staging-list release. & At $c=k_{\rm erl}$; joint-outcome rate $\lambda_Tp(\xi\mid\mathcal X)$.\\[3pt]
Prefill progress & For $(u,w,q,r)\in\mathcal P$, set $H_p\leftarrow(u,w+q,r)$ if $w+q<p(u)$. Retain an unscheduled zero-grant unfinished prefill unchanged; otherwise clear $H_p$ when all prefills finish. & Iteration completion; at most one unfinished prefill.\\[3pt]
Prefill completion & When $w+q=p(u)$: if $k=0,j=0$, $n_{\rm think}\leftarrow n_{\rm think}+1$; otherwise $\mathbf n\leftarrow\mathbf n+\mathbf e_{r,a,j,k+1}$. & Iteration completion; emits one token. Reconstruction preserves $j$.\\[3pt]
Ordinary decode & Each scheduled request in $(r,a,j,k)$ contributes $-\mathbf e_{r,a,j,k}+\mathbf e_{r,a,j',k+1}$ to $\Delta\mathbf n$, or $-\mathbf e_{r,a,j,k}$ and $+1$ to $\Delta n_{\rm think}$ on completion. & Iteration completion; probabilities $\Omega_{jj'}$ and $\omega_j$, respectively.\\[3pt]
Staging-list release & Age all surviving resident indices by $r\leftarrow\min(r+1,R_c-1)$, merging counts. Apply Equation~\ref{eq:staging-release}, appending newly eligible requests to $\mathcal Q$. Set $\mathcal P=\varnothing$, $c=0$. & After all work updates; then select the next batch, or wait for a submission if idle.\\
\end{longtable}
\endgroup

\subsection{Performance Rewards}
\label{app:rewards}
Throughput, TTFT, and request-weighted TPOT are defined through request timestamps and completion rewards~\cite{bolch2006}.

For each request $i$ we record the submission, first-token, and completion times $t_i^{\rm sub},t_i^{\rm first},t_i^{\rm end}$, and at completion we set $A_i=t_i^{\rm first}-t_i^{\rm sub}$ and $D_i=t_i^{\rm end}-t_i^{\rm first}$. The timestamps cover queueing and eligibility delays, reconstruction, and waits after eviction, and they do not alter the request lifecycle. The final output count equals $G_i=k_i^-+1$, with $k_i^-$ the progress just before completion, tokens emitted by reconstruction included. An indicator is written $\mathbf1\{\cdot\}$.

\paragraph{Completion rewards.}
At an iteration-completion jump, $\mathcal C$ holds every completing request, simultaneous completions included. The completion, multi-token completion, TTFT, and request-weighted TPOT rewards are
\begin{align}
\mathcal R_C&=|\mathcal C|,& \mathcal R_M&=\sum_{i\in\mathcal C}\mathbf1\{G_i>1\},\nonumber\\
\mathcal R_F&=\sum_{i\in\mathcal C}A_i,&
\mathcal R_P&=\sum_{\substack{i\in\mathcal C\\G_i>1}}\frac{D_i}{G_i-1}.
\end{align}
Submissions and intermediate clock-stage transitions earn zero reward. For a reward $\mathcal R_\ell$, with $\ell\in\{C,M,F,P\}$, let $J_\ell(t)$ denote the value accumulated up to time $t$. From the long-run reward rates $\rho_\ell=\lim_{t\to\infty}J_\ell(t)/t$ we obtain
\begin{equation}
X=\rho_C,\qquad
\overline{\mathrm{TTFT}}=\frac{\rho_F}{\rho_C},\qquad
\overline{\mathrm{TPOT}}=\frac{\rho_P}{\rho_M}.
\end{equation}
The expressions assume that the corresponding stable ergodic limits exist and that the denominators are positive. For a workload with no multi-token response, TPOT is undefined.

\paragraph{Reward rates and numerical evaluation.}
If the Erlang clock is finite, the expected reward rate is a function of the full execution state, and for latency rewards also of the request timestamps, collected in $\mathcal T$. Outside the last clock stage the instantaneous expected reward rate vanishes, and at $c=k_{\rm erl}$ it equals $\lambda_T\sum_\xi p(\xi\mid\mathcal X)E[\mathcal R_\ell\mid\xi,\mathcal X,\mathcal T]$. The conditional expectation averages over uniform assignments of destination counts to the requests of each cell, with timestamps held fixed. Counts alone fix the completion rewards, whereas latency rewards depend also on which request gets which destination. Averaged over time in the stationary regime, the expected reward rate equals $\rho_\ell$.

In the deterministic-duration limit the same rewards accrue at iteration boundaries spaced by the modeled durations $T$, so the long-run reward-per-time definitions apply directly.

TPOT is the per-request average of the ratio $D_i/(G_i-1)$. Dividing the total decode span by the total count of inter-token intervals gives a token-weighted mean instead.

\section{Architecture Counts and \texorpdfstring{Timing-Parameter Estimation}{Timing-Parameter Estimation}}
\label{app:timing}
The iteration-cost components, the attention work, and the graph padding are defined in Section~3.3 of the main paper. Here we add further architectural counts and the parameterization of the fit.

\subsection{Architectural Counts and Fitting Parameterization}
We write $d,d_{\rm KV},d_{\rm ff}$ for the hidden, aggregate KV-head, and feed-forward widths, $\ell$ for the layer count, and $V$ for the vocabulary size. For the supported decoder architectures, the linear-layer parameter count is
\begin{equation}P=\ell[d(d+2d_{\rm KV})+d^2+3dd_{\rm ff}].\end{equation}
Separate element sizes $b_w,b_{\rm KV}$ for weights and cache give the byte counts $W=b_wP$, $H=b_wdV$, and $C_{\rm KV}=2b_{\rm KV}\ell d_{\rm KV}$.

Timing takes the post-eviction decode counts and contexts and the prefill entries with positive grants. Every prefill/decode composition is subject to the main paper's graph-padding approximation, which alters only dense-layer arithmetic and neither consumes scheduler tokens nor advances requests. The KV-read multiplier of mixed iterations estimates the effective read cost. Cache behavior and kernel dispatch also influence the physical traffic.

The GPU specification supplies the bandwidth $B$. At the operating core clock $f$, the peak throughput is $F=F_{\rm spec}(f/f_{\rm spec})$, which rescales the dense peak specified for the configured precision at the reference clock $f_{\rm spec}$~\cite{nvidia-gpu-performance}. Both stay fixed within one serving configuration.

All times are in seconds. Three coefficients are fitted, the compute efficiency $\eta_c$, the constant residual coefficient $c_0$, and the per-decode-request residual coefficient $c_d$. As the clock frequency $f$ is given in cycles/s, $c_0$ is expressed in cycles and $c_d$ in cycles per decode request. Both are effective residual coefficients, whose time contributions at the profiling frequency $f_0$ are $c_0/f_0$ and $c_d/f_0$. For an operating clock $f$ other than $f_0$, the model keeps $c_0,c_d$ and substitutes $f$ in the main paper's residual $(c_0+c_dn_d)/f$, which assumes invariant clock-normalized residual costs.

\subsection{Probe Design and Profiling}
The 18 model--GPU pairs share one grid of candidate probe compositions, restricted for each pair to its configured KV-cache capacity. Pure-decode batches pair the request counts $1,2,4,8,16,32,64,128,256$ with the initial context lengths $128,256,512,1024,2048,4096$. Pure-prefill chunks hold $64,128,256,512,1024,2048$ tokens. Mixed batches pair decode counts of at most 128 and initial contexts of $128,512,2048$ tokens with feasible prefill chunks from that same list, and also with a chunk that fills the rest of the iteration token budget. No grid point violates the context or sequence limit. A decode or mixed composition is kept only if its planned cache occupancy stays within 90.0\% of the configured KV pool. Contexts are limited to 4096 tokens and iterations to a budget of 2048 tokens, and the sequence limit is 512. Different model--GPU pairs can therefore end up with different numbers of feasible compositions.

Every probe task measures a single batch composition and keeps 32 iterations for a pure-decode composition, eight for a pure-prefill one, and four for a mixed one. Decode measurement begins once all requests are in decode and two further iterations have been thrown away. Prefill and mixed measurements come after a warm-up repetition that is excluded.

\subsection{\texorpdfstring{Fitting}{Fitting} Objective and Optimization}
The main paper's iteration-duration equation is fitted to measured durations by minimizing a weighted sum of squared relative errors. A fitting cell collects the observations of one probe task that share prefill-token count, decode count, and cached-context decile. Per cell, the measured durations and work counts are averaged and the timing equation is evaluated at the mean counts. Every task starts with the same total weight, split equally over its cells. The weights are then rescaled inside each class so that the totals for decode, prefill, and mixed observations are 0.4749, 0.0110, and 0.5141, and finally normalized across all cells.

The optimizer starts from $\eta_c=0.75$, $c_0=(1\,\mathrm{ms})f_0$, and $c_d=(0.03\,\mathrm{ms/request})f_0$ and is constrained by $0.05\leq\eta_c\leq1$, $0\leq c_0\leq(20\,\mathrm{ms})f_0$, and $0\leq c_d\leq(1\,\mathrm{ms/request})f_0$. No additional efficiency is fitted for the memory terms, which use the specified bandwidth. The fit runs bounded trust-region reflective least squares with two-point finite differences, a linear loss, a limit of 4000 function evaluations, and tolerances of $10^{-8}$ on the function, the step, and the gradient.

Each of the 18 baseline fits uses observations from a single clock, and clock scaling is assessed separately below.

\subsection{Coefficient Estimates and Fit Diagnostics}
\label{app:fit-diagnostics}
For every model--GPU pair, Table~\ref{tab:calibration-coefficients} lists the three fitted coefficients. Its residual columns give $c_0/f_0$ and $c_d/f_0$ as times at the profiling clock $f_0$.
\begin{table}[!htbp]\centering\small
\caption{Timing coefficients fitted for all 18 model--GPU pairs. The residuals are given at the profiling clock $f_0$, which is 1200.0\,MHz on the A100, 1375.0\,MHz on the H100 SXM, and 1500.0\,MHz on the RTX 3090 Ti. Every entry is a point estimate rounded to one decimal.}\label{tab:calibration-coefficients}
\begin{tabular}{@{}llrrr@{}}\toprule
GPU & Model & \begin{tabular}{@{}r@{}}Compute efficiency\\$\eta_c$ (\%)\end{tabular} & \begin{tabular}{@{}r@{}}Constant residual\\$c_0/f_0$ (ms)\end{tabular} & \begin{tabular}{@{}r@{}}Residual per decode request\\$c_d/f_0$ ($\mu$s)\end{tabular}\\\midrule
A100 & Llama-3.2-3B & 65.8 & 3.5 & 6.5 \\
 & Llama-3-8B & 68.5 & 4.7 & 19.1 \\
 & Qwen2.5-3B & 60.5 & 4.0 & 0.0 \\
 & Qwen2.5-7B & 68.1 & 4.3 & 3.5 \\
 & DS-Coder-1.3B & 51.2 & 2.7 & 3.6 \\
 & DS-LLM-7B & 65.1 & 3.9 & 17.0 \\
\midrule
H100 SXM & Llama-3.2-3B & 63.8 & 2.4 & 4.0 \\
 & Llama-3-8B & 73.6 & 3.1 & 5.6 \\
 & Qwen2.5-3B & 59.7 & 3.0 & 0.7 \\
 & Qwen2.5-7B & 74.1 & 3.1 & 0.8 \\
 & DS-Coder-1.3B & 48.9 & 2.0 & 6.5 \\
 & DS-LLM-7B & 72.9 & 2.8 & 20.5 \\
\midrule
RTX 3090 Ti & Llama-3.2-3B & 68.8 & 3.9 & 13.6 \\
 & Llama-3-8B & 73.5 & 6.3 & 28.4 \\
 & Qwen2.5-3B & 67.9 & 4.4 & 12.6 \\
 & Qwen2.5-7B & 78.7 & 5.6 & 0.0 \\
 & DS-Coder-1.3B & 65.0 & 2.7 & 54.2 \\
 & DS-LLM-7B & 69.7 & 5.0 & 165.5 \\
\bottomrule

\end{tabular}\end{table}

\paragraph{Training-error reporting.}
The training MAPE averages the absolute percentage errors of the durations over the fitting cells, with the task and class weights of the fit. Table~5 of the main paper gives its median and range across the six models on each GPU. The 18 fits all converged.

\paragraph{Clock-transfer probes.}
For the clock ablation in Section~4.5 of the main paper, one A100 is probed at 1200, 900, and 750\,MHz. The probes at the lower clocks supply the refitted references, and the transferred predictions retain the parameters fitted at 1200\,MHz. When each probe is refitted on its own, $c_0$ at 900 and 750\,MHz comes out at 0.92 and 0.98 times its 1200\,MHz value, against the 0.75 and 0.63 that a constant fixed in seconds would produce. Each probe measures a reference composition, 16 decode requests with 512-token contexts, at its start, at its end, and twice in between. These four mean iteration durations may differ by at most 2\%. At 750\,MHz the spread is 2.6\%, entirely due to the first measurement, and the remaining three agree to within 0.2\%. Dropping every measurement that precedes the second reference measurement brings the ratio to 0.94.

\paragraph{Measurement boundary.}
The timing expression contains vocabulary projection, whereas the profiled interval leaves out both vocabulary projection and sampling. Because of this mismatch the fitted residual cannot be read as a physical overhead measured independently.

\FloatBarrier
\section{Trace-Based Workload Construction}
\label{app:workload-construction}
The serving results use a joint length law, which we fit to the observed prompt/output pairs.

\subsection{Joint Prompt and Generation Law}

The fitted workload law consists of prompt lengths $a$ with finite support, the transient generation matrix $\Omega$, and the joint entry probabilities $\pi_{a,j}$. Let $\boldsymbol\omega=\mathbf1-\Omega\mathbf1$ and $\boldsymbol\pi_{a,+}=(\pi_{a,1},\ldots,\pi_{a,L_g})$. Without eviction, the length law is
\begin{align}
\Pr(a,g=1)&=\pi_{a,0},\nonumber\\
\Pr(a,g=1+m)&=\boldsymbol\pi_{a,+}\Omega^{m-1}\boldsymbol\omega,\qquad m\geq1.
\end{align}
All entries of $\pi$ are nonnegative, and together they sum to one. The matrix $\Omega$ is nonnegative and substochastic (no row sums to more than one), and absorption can be reached from every reachable phase. Absorption of the fitted finite chains is therefore almost sure. Empirical averages over the observed pairs are written $\widehat E[\cdot]$, and $\widehat p_1$ is the probability of $g=1$ among those pairs.

\subsection{Prompt Approximation and Conditional Generation}
\paragraph{Prompt classes.}
Lloyd clustering in one dimension~\cite{lloyd1982}, started from equally spaced quantiles, yields centroids $z_i$ with probabilities $\rho_i^{(0)}$. Before rounding, the centroids are rescaled about the empirical mean $\bar a$ so that their variance equals the empirical prompt variance. The scale factor comes from their $\rho_i^{(0)}$-weighted variance and is set to one if that variance vanishes.

Each representative is rounded to a positive integer length $a_i$ that is no exact multiple of the KV block size, and a choice that makes two classes share a representative is rejected. Probabilities are then recomputed by minimizing $\sum_i(\rho_i-\rho_i^{(0)})^2$ under $0\leq\rho_i\leq1$, $\sum_i\rho_i=1$, and $\sum_i\rho_i a_i=\widehat E[a]$, with the further constraint $\sum_i\rho_i a_i^2=\widehat E[a^2]$ when there are at least three classes. Should the numerical residual of the constraints be above tolerance, bounded least squares approximates the normalization and the first moment, and the weights are normalized afterwards. All three fitted laws match the prompt moments exactly.

Differences between distributions are measured by the first Wasserstein distance $W_1$, the area between their two cumulative distribution functions. The number of classes $\ell_p$ is the least value in $\{1,2,3,4,6,8,12,16\}$ at which the Wasserstein distance from the class law to the observed prompt lengths stays within $0.1\,\widehat E[a]$. A single class is set at the mean prompt length, rounded to avoid exact multiples of the KV block size. Every observed prompt is assigned to the class whose representative is nearest, and both fitting and scoring use that assignment.

\paragraph{Generation phases.}
In this length distribution, $D=g-1$ counts the output tokens after the first. Given $D>0$, the output is drawn from a mixture of $B$ branches. Each branch $b$ chains $k_b$ geometric phases with a common success probability $q_b=k_b/\mu_b$ and has mean length $\mu_b\geq k_b$, which gives
\begin{equation}
 \Pr(D=d\mid b)=\binom{d-1}{k_b-1}q_b^{k_b}(1-q_b)^{d-k_b},\qquad d\geq k_b.
\end{equation}
The discrete expectation-maximization updates given below mirror hyper-Erlang fitting in continuous time~\cite{thummler-tdsc06}. The matrix $\Omega$ is built from the geometric self-loops and the phase-to-phase transitions and has $L_g=\sum_b k_b$ phases. Entry probability is nonzero only in the first phase of each branch. All prompt classes share the branch means and the phase counts.

\paragraph{Prompt-dependent weights.}
Adjacent prompt classes may share one output mixture. A Lloyd clustering of the class representatives, with each prompt class $a$ weighted by its fitted probability $\rho_a$, partitions the $\ell_p$ classes into $\ell_c\leq\ell_p$ groups of adjacent classes. The classes of a group $c$ have common branch weights $w_{c,b}$ and a common single-token probability $\widehat p_{1,c}$, which is the share of the group's requests observed with $g=1$. A class $a$ of group $c$ then has
\begin{equation}
 \pi_{a,0}=\rho_a\,\widehat p_{1,c},\qquad
 \pi_{a,e_b}=\rho_a\,(1-\widehat p_{1,c})\,w_{c,b},
\end{equation}
where $e_b$ denotes the first phase of branch $b$. With a single group, prompt and output lengths are independent, and with one group per class every class has an output mixture of its own.

\subsection{Likelihood Fitting and Structure Selection}
With the phase counts held fixed, expectation maximization estimates $\mu_b$ and $w_{c,b}$ from the observed counts per (group, $D$). The responsibilities are the posterior branch probabilities. Each update sets $\mu_b$ to the mean of $D$ weighted by responsibility and $w_{c,b}$ to the fraction of group $c$'s responsibility that falls on branch $b$. Output lengths up to four KV blocks are counted exactly. Longer ones fall into bins one block wide, and each bin is represented by its mean length.

Phase counts are chosen on the ladder $\{1,2,3,5,8,13,21,34,55,89,144,233\}$ by a search in which every step raises the training likelihood. From $k_b=1$, each step advances the one branch whose next rung gives the largest likelihood gain, considering only moves under which $L_g$ respects a configured cap on the total phase count. The search ends when no move raises the likelihood.

\paragraph{Held-out score.}
The search runs over a grid of group counts $\ell_c\in\{1,2,3,4,6,8,12,16\}$ with $\ell_c\leq\ell_p$ and branch counts $B\in\{1,2,3,4\}$. At each grid point the pairs are split into halves by five seeded random splits, and each half is fitted and scored on the other, for ten fits in total. With $F_{g\mid a}$ the fitted conditional distribution of output length, $\widehat\rho_a$ and $\widehat F_{g\mid a}$ the class shares and the empirical distributions of the held-out half, and $\widehat E[g]$ the mean over that half, the score is
\begin{equation}
 S=\frac{1}{\widehat E[g]}\sum_{a}\widehat\rho_a\,W_1\bigl(F_{g\mid a},\widehat F_{g\mid a}\bigr).
\end{equation}
The Wasserstein distance $W_1$ takes the entire fitted tail into account, and all candidates are scored on the same $\ell_p$ prompt classes.

\paragraph{Admissibility and final fit.}
Admissibility is assessed after a candidate is fitted on all pairs. A candidate is admissible only if expectation maximization converged, completion can be reached from every reachable phase, and its duration-weighted mean KV footprint lies within 1\% of the observed value $\sum_i g_i(a_i+(g_i+1)/2)/\sum_i g_i+\sigma$. Among the admissible candidates, the one-standard-error rule~\cite{hastie-esl09} keeps those whose mean score exceeds the minimum by at most one standard error. Of these it picks the one with the fewest phases (the median over the ten fits), with ties broken by group count and then by branch count.

Without an admissible candidate, the footprint tolerance is loosened to 2.5\%, and if none is admissible even then, one common factor rescales all branch means to restore the footprint exactly. The final law comes from refitting the chosen structure on all pairs with at most 12 phases. The fit draws only on the observed length pairs, never on serving measurements.

\paragraph{Workload sampling.}
ServeCast draws the prompt and the entry phase jointly and generates the output through phase transitions. Vidur, like the hardware clients, replays the prompt/output pairs of the trace slice. Reconstruction can change the output lengths of ServeCast, as Section~\ref{app:replay} describes.

\section{Experimental Traces and Cache Capacities}
\label{app:experiment-config}
Each campaign is defined by its model, GPU, and trace, and a configuration adds the client population $N$. In each campaign, the six values of $N$ are rounded from $c$, our workload-based estimate of how many requests fit in the KV cache, with
\[
N/c\in\{0.271,\;0.406,\;0.541,\;0.812,\;1.082,\;1.623\}.
\]
For example, A100/Llama-3-8B/BurstGPT gets $N=44,66,88,132,176,264$.

Joint prompt/output lengths come from the traces, and submission times from the closed clients~\cite{burstgpt-kdd25,splitwise-isca24,servegen-nsdi26,acdc-sosp26,schroeder-openclosed06}. We keep requests with positive lengths and a prompt plus output of at most 4096 tokens. Longer requests are removed whole, without truncation.

Sampling BurstGPT at a fixed stride gives a slice of 50,000 successful requests. The 17,754 requests of the Azure slice come from the conversation trace in the Azure LLM Inference Dataset 2023, which accompanied Splitwise~\cite{splitwise-isca24}. The ServeGen slice holds 44,990 text requests drawn from offline trace D.

Table~\ref{tab:cache-config} lists the allocated KV-cache capacity $C_{\rm alloc}$ of each model--GPU pair. ServeCast holds back one 16-token block and works with $C=C_{\rm alloc}-B_{\rm blk}$.

\begin{table}[htbp]\centering\small
\caption{Allocated KV-cache tokens per model and GPU. The usable capacity in ServeCast is 16 tokens lower.}\label{tab:cache-config}
\begin{tabular}{@{}lrrr@{}}\toprule
Model & A100 & H100 SXM & RTX 3090 Ti\\\midrule
Llama-3.2-3B & 159440 & 159440 & 129984\\
Llama-3-8B & 159440 & 159440 & 43904\\
Qwen2.5-3B & 159440 & 159440 & 159440\\
Qwen2.5-7B & 159440 & 159440 & 114112\\
DS-Coder-1.3B & 159440 & 159440 & 93856\\
DS-LLM-7B & 45920 & 45920 & 16032\\\bottomrule
\end{tabular}\end{table}

\section{Supplementary Results}
\label{app:complete-results}

\subsection{Paired Prediction Time}
\label{app:campaign-runtime}
Table~\ref{tab:campaign-runtime} breaks down the grouped prediction times of Table~5 in the main paper. For ServeCast the wall time covers checkpoint setup, the simulated execution, and the statistical calculations. For Vidur it covers initializing the simulator and the timing predictor, executing, and extracting metrics. Neither includes the interpreter launch or the earlier preparation of inputs. Vidur works from its cached timing predictor, and the RTX Azure timings leave out the one-time cost of training it. With a single execution per method and campaign, these timings do not capture how repeated runs vary.

\begingroup\small
\renewcommand{\arraystretch}{1.05}
\begin{longtable}{@{}lllrrr@{}}
\caption{Wall time of prediction per configuration, in seconds, as the arithmetic mean over the six client populations of a model--GPU--trace campaign. The ratio divides the mean of Vidur by that of ServeCast over identical configurations.}\label{tab:campaign-runtime}\\
\toprule
GPU & Model & Trace & ServeCast & Vidur & Ratio\\\midrule
\endfirsthead
\multicolumn{6}{l}{\tablename~\thetable\ continued}\\\toprule
GPU & Model & Trace & ServeCast & Vidur & Ratio\\\midrule
\endhead
\midrule\multicolumn{6}{r}{Continued on next page}\\\endfoot
\bottomrule\endlastfoot
A100 & Llama-3.2-3B & BurstGPT & 4.0 & 16.3 & 4.1\\
A100 & Llama-3.2-3B & Azure & 2.2 & 5.9 & 2.7\\
A100 & Llama-3.2-3B & ServeGen & 13.4 & 42.5 & 3.2\\
A100 & Llama-3-8B & BurstGPT & 13.2 & 76.6 & 5.8\\
A100 & Llama-3-8B & Azure & 3.6 & 3.3 & 0.9\\
A100 & Llama-3-8B & ServeGen & 11.7 & 41.9 & 3.6\\
A100 & Qwen2.5-3B & BurstGPT & 3.8 & 11.8 & 3.1\\
A100 & Qwen2.5-3B & Azure & 2.6 & 3.1 & 1.2\\
A100 & Qwen2.5-3B & ServeGen & 14.0 & 45.9 & 3.3\\
A100 & Qwen2.5-7B & BurstGPT & 12.5 & 75.7 & 6.1\\
A100 & Qwen2.5-7B & Azure & 3.6 & 2.9 & 0.8\\
A100 & Qwen2.5-7B & ServeGen & 13.0 & 52.7 & 4.1\\
A100 & DS-Coder-1.3B & BurstGPT & 4.1 & 16.2 & 3.9\\
A100 & DS-Coder-1.3B & Azure & 1.5 & 5.9 & 4.0\\
A100 & DS-Coder-1.3B & ServeGen & 12.6 & 50.6 & 4.0\\
A100 & DS-LLM-7B & BurstGPT & 9.9 & 13.2 & 1.3\\
A100 & DS-LLM-7B & Azure & 2.1 & 4.5 & 2.2\\
A100 & DS-LLM-7B & ServeGen & 88.9 & 53.1 & 0.6\\
H100 SXM & Llama-3.2-3B & BurstGPT & 5.3 & 18.4 & 3.5\\
H100 SXM & Llama-3.2-3B & Azure & 3.6 & 16.6 & 4.6\\
H100 SXM & Llama-3.2-3B & ServeGen & 13.5 & 48.1 & 3.6\\
H100 SXM & Llama-3-8B & BurstGPT & 4.2 & 12.8 & 3.0\\
H100 SXM & Llama-3-8B & Azure & 2.4 & 10.6 & 4.5\\
H100 SXM & Llama-3-8B & ServeGen & 13.8 & 41.2 & 3.0\\
H100 SXM & Qwen2.5-3B & BurstGPT & 6.1 & 18.8 & 3.1\\
H100 SXM & Qwen2.5-3B & Azure & 2.9 & 5.4 & 1.9\\
H100 SXM & Qwen2.5-3B & ServeGen & 16.3 & 65.9 & 4.0\\
H100 SXM & Qwen2.5-7B & BurstGPT & 3.4 & 12.0 & 3.5\\
H100 SXM & Qwen2.5-7B & Azure & 2.9 & 7.8 & 2.7\\
H100 SXM & Qwen2.5-7B & ServeGen & 15.4 & 70.5 & 4.6\\
H100 SXM & DS-Coder-1.3B & BurstGPT & 11.9 & 16.6 & 1.4\\
H100 SXM & DS-Coder-1.3B & Azure & 2.7 & 10.4 & 3.9\\
H100 SXM & DS-Coder-1.3B & ServeGen & 13.8 & 73.4 & 5.3\\
H100 SXM & DS-LLM-7B & BurstGPT & 9.2 & 14.6 & 1.6\\
H100 SXM & DS-LLM-7B & Azure & 3.4 & 7.3 & 2.2\\
H100 SXM & DS-LLM-7B & ServeGen & 68.5 & 43.7 & 0.6\\
RTX 3090 Ti & Llama-3.2-3B & BurstGPT & 4.5 & 15.2 & 3.4\\
RTX 3090 Ti & Llama-3.2-3B & Azure & 3.9 & 4.5 & 1.2\\
RTX 3090 Ti & Llama-3.2-3B & ServeGen & 10.7 & 30.8 & 2.9\\
RTX 3090 Ti & Llama-3-8B & BurstGPT & 4.3 & 8.5 & 2.0\\
RTX 3090 Ti & Llama-3-8B & Azure & 1.9 & 2.1 & 1.1\\
RTX 3090 Ti & Llama-3-8B & ServeGen & 89.0 & 46.9 & 0.5\\
RTX 3090 Ti & Qwen2.5-3B & BurstGPT & 3.8 & 16.7 & 4.4\\
RTX 3090 Ti & Qwen2.5-3B & Azure & 4.3 & 2.8 & 0.6\\
RTX 3090 Ti & Qwen2.5-3B & ServeGen & 12.1 & 54.6 & 4.5\\
RTX 3090 Ti & Qwen2.5-7B & BurstGPT & 5.3 & 13.9 & 2.6\\
RTX 3090 Ti & Qwen2.5-7B & Azure & 2.6 & 2.3 & 0.9\\
RTX 3090 Ti & Qwen2.5-7B & ServeGen & 9.3 & 41.5 & 4.5\\
RTX 3090 Ti & DS-Coder-1.3B & BurstGPT & 7.5 & 15.9 & 2.1\\
RTX 3090 Ti & DS-Coder-1.3B & Azure & 2.1 & 3.6 & 1.7\\
RTX 3090 Ti & DS-Coder-1.3B & ServeGen & 19.9 & 31.7 & 1.6\\
RTX 3090 Ti & DS-LLM-7B & BurstGPT & 9.3 & 12.5 & 1.3\\
RTX 3090 Ti & DS-LLM-7B & Azure & 2.0 & 3.9 & 2.0\\
RTX 3090 Ti & DS-LLM-7B & ServeGen & 97.2 & 74.9 & 0.8\\

\end{longtable}
\endgroup

\subsection{Transfer Accuracy by Metric}
\label{app:transfer-metrics}
Table~4 of the main paper gives the TTFT matrix, and Tables~\ref{tab:rq3-throughput} and~\ref{tab:rq3-tpot} give the errors in throughput and in request-weighted TPOT for those same transfers.

\begin{table}[!htbp]\centering\small
\caption{ServeCast throughput MAPE (\%) when transferred between models on the same GPU. Values are averaged first over each trace's six client populations and then with equal weight across the three traces. Rows give the source model that was profiled and fitted, grouped by GPU, and each column is the target model. The light blue diagonal is the target-profiled reference of RQ1. In each GPU group, the last two rows set ServeCast, averaged over the five other source models, against target-profiled Vidur, and green (ServeCast) or amber (Vidur) shows which error is lower before rounding. In the final column, source rows average over the five off-diagonal targets and comparison rows over all six targets.}\label{tab:rq3-throughput}
\setlength{\tabcolsep}{3.5pt}
\begin{tabular}{@{}llcccccc@{\hspace{8pt}}c}\toprule
 & & \multicolumn{6}{c}{Predicted for} & \\\cmidrule(lr){3-8}
GPU & Profiled on & Llama-3.2-3B & Llama-3-8B & Qwen2.5-3B & Qwen2.5-7B & DS-Coder-1.3B & DS-LLM-7B & \begin{tabular}{@{}c@{}}Mean over\\models\end{tabular}\\\midrule
\multirow{8}{*}{\rotatebox[origin=c]{90}{A100}} & Llama-3.2-3B & \arxivdiag{4.9} & 7.7 & 7.6 & 5.9 & 4.4 & 3.3 & 5.8 \\
 & Llama-3-8B & 3.3 & \arxivdiag{4.1} & 4.8 & 4.1 & 7.3 & 3.2 & 4.5 \\
 & Qwen2.5-3B & 3.8 & 8.4 & \arxivdiag{5.2} & 8.4 & 4.4 & 3.0 & 5.6 \\
 & Qwen2.5-7B & 4.5 & 7.7 & 7.1 & \arxivdiag{4.5} & 4.8 & 2.4 & 5.3 \\
 & DS-Coder-1.3B & 7.5 & 13.8 & 7.3 & 14.5 & \arxivdiag{2.9} & 5.7 & 9.8 \\
 & DS-LLM-7B & 2.3 & 6.3 & 3.6 & 5.6 & 4.4 & \arxivdiag{2.3} & 4.4 \\
\cmidrule(l){2-9}
 & ServeCast, transferred & \winS{4.3} & \winS{8.8} & \winS{6.0} & \winS{7.7} & 5.1 & \winS{3.5} & \winS{5.9} \\
 & Vidur, profiled & 10.5 & 13.5 & 17.7 & 15.0 & \winV{3.7} & 3.9 & 10.7 \\
\midrule
\multirow{8}{*}{\rotatebox[origin=c]{90}{H100 SXM}} & Llama-3.2-3B & \arxivdiag{7.9} & 7.7 & 11.1 & 5.5 & 5.3 & 8.4 & 7.6 \\
 & Llama-3-8B & 4.4 & \arxivdiag{6.6} & 6.4 & 4.4 & 4.4 & 5.3 & 5.0 \\
 & Qwen2.5-3B & 4.5 & 5.1 & \arxivdiag{5.7} & 6.1 & 4.1 & 4.2 & 4.8 \\
 & Qwen2.5-7B & 7.1 & 9.3 & 9.3 & \arxivdiag{6.5} & 5.5 & 6.5 & 7.6 \\
 & DS-Coder-1.3B & 6.3 & 10.3 & 7.4 & 11.9 & \arxivdiag{4.5} & 7.6 & 8.7 \\
 & DS-LLM-7B & 3.7 & 2.3 & 4.4 & 2.9 & 7.1 & \arxivdiag{4.5} & 4.1 \\
\cmidrule(l){2-9}
 & ServeCast, transferred & 5.2 & 6.9 & \winS{7.7} & \winS{6.2} & \winS{5.3} & 6.4 & \winS{6.3} \\
 & Vidur, profiled & \winV{5.1} & \winV{5.3} & 14.2 & 8.2 & 7.1 & \winV{5.0} & 7.5 \\
\midrule
\multirow{8}{*}{\rotatebox[origin=c]{90}{RTX 3090 Ti}} & Llama-3.2-3B & \arxivdiag{11.4} & 11.6 & 12.7 & 9.1 & 18.9 & 11.0 & 12.7 \\
 & Llama-3-8B & 7.1 & \arxivdiag{7.3} & 7.7 & 5.6 & 8.8 & 4.6 & 6.8 \\
 & Qwen2.5-3B & 9.6 & 10.1 & \arxivdiag{11.1} & 9.5 & 16.6 & 8.8 & 10.9 \\
 & Qwen2.5-7B & 17.3 & 13.8 & 20.4 & \arxivdiag{6.5} & 19.3 & 8.5 & 15.9 \\
 & DS-Coder-1.3B & 6.8 & 11.3 & 7.3 & 12.1 & \arxivdiag{14.7} & 14.8 & 10.5 \\
 & DS-LLM-7B & 9.6 & 5.0 & 11.9 & 13.6 & 8.3 & \arxivdiag{4.0} & 9.7 \\
\cmidrule(l){2-9}
 & ServeCast, transferred & 10.1 & 10.4 & \winS{12.0} & 10.0 & 14.4 & 9.6 & 11.1 \\
 & Vidur, profiled & \winV{10.0} & \winV{3.7} & 20.3 & \winV{8.1} & \winV{4.1} & \winV{5.4} & \winV{8.6} \\
\bottomrule

\end{tabular}\end{table}

\begin{table}[!htbp]\centering\small
\caption{ServeCast request-weighted TPOT MAPE (\%) when transferred between models on the same GPU. Values are averaged first over each trace's six client populations and then with equal weight across the three traces. Rows give the source model that was profiled and fitted, grouped by GPU, and each column is the target model. The light blue diagonal is the target-profiled reference of RQ1. In each GPU group, the last two rows set ServeCast, averaged over the five other source models, against target-profiled Vidur, and green (ServeCast) or amber (Vidur) shows which error is lower before rounding. In the final column, source rows average over the five off-diagonal targets and comparison rows over all six targets.}\label{tab:rq3-tpot}
\setlength{\tabcolsep}{3.5pt}
\begin{tabular}{@{}llcccccc@{\hspace{8pt}}c}\toprule
 & & \multicolumn{6}{c}{Predicted for} & \\\cmidrule(lr){3-8}
GPU & Profiled on & Llama-3.2-3B & Llama-3-8B & Qwen2.5-3B & Qwen2.5-7B & DS-Coder-1.3B & DS-LLM-7B & \begin{tabular}{@{}c@{}}Mean over\\models\end{tabular}\\\midrule
\multirow{8}{*}{\rotatebox[origin=c]{90}{A100}} & Llama-3.2-3B & \arxivdiag{6.1} & 7.9 & 8.3 & 7.1 & 5.1 & 3.8 & 6.4 \\
 & Llama-3-8B & 4.1 & \arxivdiag{5.5} & 6.0 & 5.3 & 8.2 & 2.3 & 5.2 \\
 & Qwen2.5-3B & 5.4 & 9.0 & \arxivdiag{6.1} & 10.3 & 5.3 & 2.4 & 6.5 \\
 & Qwen2.5-7B & 5.1 & 7.7 & 7.5 & \arxivdiag{6.2} & 5.7 & 2.5 & 5.7 \\
 & DS-Coder-1.3B & 8.9 & 15.2 & 8.8 & 17.5 & \arxivdiag{4.0} & 5.2 & 11.1 \\
 & DS-LLM-7B & 3.4 & 7.0 & 4.4 & 7.0 & 4.4 & \arxivdiag{1.9} & 5.2 \\
\cmidrule(l){2-9}
 & ServeCast, transferred & \winS{5.4} & \winS{9.4} & \winS{7.0} & \winS{9.4} & 5.7 & \winS{3.2} & \winS{6.7} \\
 & Vidur, profiled & 9.8 & 11.3 & 16.2 & 12.3 & \winV{5.2} & 3.8 & 9.8 \\
\midrule
\multirow{8}{*}{\rotatebox[origin=c]{90}{H100 SXM}} & Llama-3.2-3B & \arxivdiag{10.4} & 9.2 & 13.8 & 6.3 & 8.8 & 11.6 & 9.9 \\
 & Llama-3-8B & 6.3 & \arxivdiag{8.2} & 8.3 & 5.2 & 6.3 & 7.0 & 6.6 \\
 & Qwen2.5-3B & 5.9 & 6.3 & \arxivdiag{7.5} & 8.1 & 5.8 & 5.9 & 6.4 \\
 & Qwen2.5-7B & 9.4 & 10.8 & 11.8 & \arxivdiag{7.3} & 7.6 & 8.3 & 9.6 \\
 & DS-Coder-1.3B & 7.5 & 12.3 & 9.0 & 15.5 & \arxivdiag{7.5} & 10.2 & 10.9 \\
 & DS-LLM-7B & 3.6 & 4.3 & 5.6 & 2.8 & 9.0 & \arxivdiag{6.5} & 5.0 \\
\cmidrule(l){2-9}
 & ServeCast, transferred & 6.5 & 8.6 & \winS{9.7} & \winS{7.6} & \winS{7.5} & 8.6 & \winS{8.1} \\
 & Vidur, profiled & \winV{5.4} & \winV{4.8} & 15.0 & 8.0 & 10.2 & \winV{6.6} & 8.3 \\
\midrule
\multirow{8}{*}{\rotatebox[origin=c]{90}{RTX 3090 Ti}} & Llama-3.2-3B & \arxivdiag{11.7} & 11.4 & 12.3 & 9.0 & 18.7 & 12.3 & 12.7 \\
 & Llama-3-8B & 8.4 & \arxivdiag{8.3} & 8.9 & 5.4 & 9.9 & 5.0 & 7.5 \\
 & Qwen2.5-3B & 10.0 & 10.2 & \arxivdiag{10.2} & 10.1 & 16.3 & 10.5 & 11.4 \\
 & Qwen2.5-7B & 16.7 & 14.2 & 18.4 & \arxivdiag{6.3} & 18.5 & 9.8 & 15.5 \\
 & DS-Coder-1.3B & 6.4 & 11.2 & 6.2 & 12.9 & \arxivdiag{15.5} & 14.1 & 10.2 \\
 & DS-LLM-7B & 9.0 & 5.2 & 13.3 & 15.3 & 8.5 & \arxivdiag{5.1} & 10.3 \\
\cmidrule(l){2-9}
 & ServeCast, transferred & 10.1 & 10.4 & \winS{11.8} & 10.5 & 14.4 & 10.3 & 11.3 \\
 & Vidur, profiled & \winV{9.0} & \winV{4.0} & 16.5 & \winV{6.9} & \winV{4.9} & \winV{4.8} & \winV{7.7} \\
\bottomrule

\end{tabular}\end{table}

\subsection{Accuracy by Load}
\label{app:load-bands}
In Table~\ref{tab:load-bands}, the RQ1 errors are grouped by load. Each campaign's six client populations are split into three bands of two, ordered from lightest to heaviest, and the populations of the high band exceed the estimated KV-cache residency $c$. In the high band, ServeCast's errors remain on a par with those of the lighter bands, and its error there is below Vidur's on all three metrics.

\begin{table}[htbp]\centering\small
\caption{RQ1 MAPE (\%) per load band, pooled across all model--GPU--trace combinations, with the lower error shaded.}\label{tab:load-bands}
\begin{tabular}{@{}l rrr@{\hspace{9pt}}rrr@{}}\toprule
 & \multicolumn{3}{c}{ServeCast} & \multicolumn{3}{c}{Vidur}\\\cmidrule(lr){2-4}\cmidrule(l){5-7}
Load band & Thr. & TTFT & TPOT & Thr. & TTFT & TPOT\\\midrule
Low ($N/c=0.27$, $0.41$) & \winS{6.0} & \winS{11.4} & 7.1 & 6.4 & 27.6 & \winV{6.6} \\
Mid ($N/c=0.54$, $0.81$) & \winS{7.6} & \winS{12.6} & \winS{8.0} & 10.0 & 32.2 & 9.5 \\
High ($N/c=1.08$, $1.62$) & \winS{5.6} & \winS{11.7} & \winS{7.4} & 10.6 & 14.9 & 9.9 \\
\bottomrule

\end{tabular}\end{table}

\end{document}